\documentclass[final,aps,prb,10pt,amsmath,amssymb]{revtex4-2}
\usepackage{amssymb,amsfonts,esint}
\usepackage{array}
\usepackage{booktabs,tabularx,array,threeparttable}
\usepackage[most]{tcolorbox}
\usepackage[table]{xcolor}
\usepackage{graphicx}
\graphicspath{{Paper/figures/}}
\usepackage{dcolumn}
\usepackage{bm}
\usepackage{tikz}
\usepackage{mathtools}
\usepackage{extarrows}
\usepackage{quantikz}
\usepackage{svg}
\usepackage{algorithm2e}
\usepackage{bbold}
\usepackage{relsize}
\usepackage{amsthm}
\usepackage{comment}
\usepackage{braket}

\usepackage{hyperref}
\usepackage{cleveref}
\hypersetup{
    colorlinks=true, 
    linkcolor=blue,  
    citecolor=blue, 
    urlcolor=blue,   
    filecolor=magenta 
}

\usetikzlibrary{calc}

\usepackage[colorinlistoftodos, textsize=small, color=orange!50]{todonotes}

\newcommand{\dg}{\dagger}

\newcommand{\mbf}{\mathbf}
\newcommand{\btheta}{\boldsymbol{\theta}}
\newcommand{\nn}{\nonumber}

\newcommand{\sign}{\text{sign}}
\newcommand{\PREP}{\text{PREP}}
\newcommand{\SEL}{\text{SEL}}
\newcolumntype{L}[1]{>{\raggedright\arraybackslash}p{#1}}
\newcolumntype{C}[1]{>{\centering\arraybackslash}p{#1}}
\tikzset{
    gridstyle/.style={
        step=0.7cm,
        color=gray!40, 
        line width=0.4pt
    },
    error/.style={
        -stealth,         
        blue!80!black,    
        line width=1.2pt, 
        rounded corners=1.5pt 
    },
    labelstyle/.style={
        anchor=west,
        font=\small,
        fill=white,       
        inner sep=1.5pt   
    }
}

\begin{document}
\usetikzlibrary{arrows.meta,positioning}
\preprint{APS/123-QED}

\title{Fault tolerant computation of the static structure factor and finite size effects}

\author{Rishabh Bhardwaj}
\email{rbhardwaj@lanl.gov}
\affiliation{Computing and Artificial Intelligence Division (CAI-3), Los Alamos National Laboratory, Los Alamos, 87545, US}
\affiliation{Center for Quantum Computing (CQC), Los Alamos National Laboratory, Los Alamos, 87545, US}

\author{Alexander Reed Mu\~{n}oz}
\email{alexmunoz@lanl.gov}
\affiliation{Theoretical Division, Los Alamos National Laboratory, Los Alamos, 87545, US}

\author{Travis E. Jones}
\email{tejones@lanl.gov}
\affiliation{Theoretical Division, Los Alamos National Laboratory, Los Alamos, 87545, US} 

\author{John Golden}
\email{golden@lanl.gov}
\affiliation{Computing and Artificial Intelligence Division (CAI-3), Los Alamos National Laboratory, Los Alamos, 87545, US}
\affiliation{Center for Quantum Computing (CQC), Los Alamos National Laboratory, Los Alamos, 87545, US}
\date{\today}


\begin{abstract}
Fault-tolerant quantum algorithms offer a promising pathway for estimating the ground-state energies of periodic materials that are beyond the practical reach of classical electronic-structure methods. A remaining challenge is finite-size mitigation: quantum algorithms evaluate a finite supercell or finite Brillouin-zone mesh, while materials properties are defined in the thermodynamic limit. In this work we develop a quantum post-processing strategy for the leading two-body finite-size correction. After one-body shell effects are reduced by twist averaging, the dominant residual error is controlled by long-wavelength density fluctuations, which are encoded in the small-momentum static structure factor $S(\mbf q)$. We formulate the corresponding operator in a Bloch-orbital basis, construct its block encoding through the density operator, and estimate its ground-state expectation value using an amplified Hadamard test. We also introduce adaptive global and local binary search procedures for identifying the infrared fitting window used to reconstruct the two-body finite size error correction. The resulting cost remains subleading relative to the main ground-state energy estimation routine: the structure-factor correction has leading $\mathcal{O}(N_bN_k)^3$ dependence on the Bloch-orbital basis size, avoids the large plane-wave prefactor of full Hamiltonian simulation, and requires only $\mathcal{O}(N_bN_k)$ logical qubits. This provides a fault-tolerant alternative to down-sampling, replacing repeated energy calculations on larger cells with targeted measurements of the infrared density correlations that control the finite-size effects.
\end{abstract}

\maketitle


\section{Introduction}

Computing ground-state energies of interacting electrons is among the most compelling applications of fault-tolerant quantum computers~\cite{Lee_2023,Gundlach_2025,Genin_2026,Cao_2019,Schleich_2025}. Over the past decade, quantum algorithms for electronic structure have evolved into increasingly practical workflows built on block encodings, qubitization, and quantum phase estimation (QPE)~\cite{Babbush_2018,Lee_2021,Low_2019,Rubin_bloch_2023,ivanov_2024_paw,Bhardwaj:2026koe}. For extended periodic systems, these methods offer a route to materials regimes where classical approximations struggle, including correlated metals, transition-metal oxides, warm dense matter, and other systems in which long-range screening, strong correlation, and finite-size convergence must be controlled simultaneously.
\\

A central obstacle in applying these algorithms to solids is that a quantum simulation does not give the ground state energy in the thermodynamic limit, $E_{\infty}$ directly. One instead computes the energy of a finite supercell, or equivalently a discrete Brillouin-zone (BZ) mesh, and then removes the resulting finite-size error in classical post-processing. This error has two principal components. The first is a one-body shell error arising from the discrete sampling of single-particle states. This error is especially pronounced in metals, where an accurate description of the BZ is required because the electronic structure is highly sensitive to states near the Fermi surface. The second is a two-body error arising from the distortion of long-wavelength Coulomb correlations in a finite periodic cell. In classical and quantum Monte Carlo (QMC) calculations, the one-body component is commonly reduced by \emph{twist averaging}, first introduced in Refs.~\cite{Poilblanc_1991_twisted,Lin_2001}. The two-body contribution is commonly addressed using one of two common strategies: down-sampling over a hierarchy of cell sizes or $k$-meshes~\cite{Shimazaki_2009,Shukri_2016,Taheridehkordi_2023,ivanov_2024_paw} and  corrections based on the small-momentum behavior of the static structure factor (SSF)~\cite{Holzmann_2016}. Down-sampling is more systematic, but it is expensive: it requires repeating the underlying energy calculation on progressively larger supercells or denser $k$-meshes. Starting from the coarsest calculation, one successively adds finite-size increments from neighboring levels of the hierarchy to estimate the thermodynamic-limit energy. If the largest required cubic mesh is size $L$, this procedure requires $2L+1$ separate energy evaluations. In the fault-tolerant setting this overhead is especially costly, because ground-state energy estimation is already the dominant quantum primitive.
\begin{figure}[t]
        \centering
        \includegraphics[
    width=1\linewidth,
    height=0.33\textheight
]{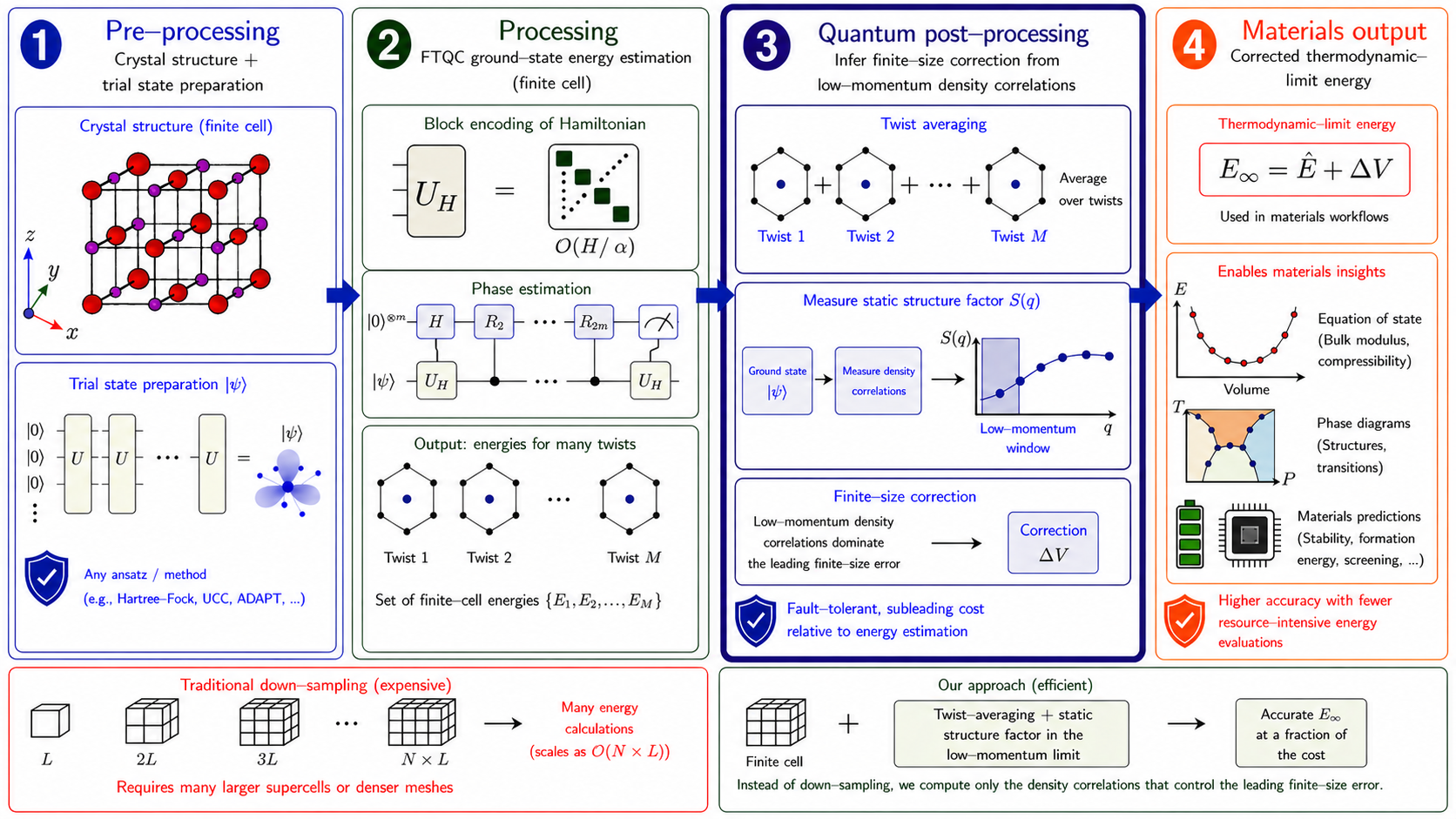}
        \caption{\small 
Schematic overview of finite-size mitigation as a quantum post-processing task of the quantum workflow of $E_{\infty}$ estimation. 
A finite periodic cell and trial state are supplied as input to a fault-tolerant ground-state energy-estimation routine, which produces finite-cell energies for different twists. 
Rather than reaching the thermodynamic limit by repeating the full energy calculation on a hierarchy of larger supercells or denser meshes, the post-processing stage combines twist averaging with targeted measurements of the small-momentum SSF. 
These long-wavelength density correlations determine the leading residual two-body finite-size correction, yielding the corrected thermodynamic-limit estimate. 
The corrected energy can then be used in downstream materials workflows such as equations of state, phase diagrams, and materials-property prediction.
\normalsize}
        \label{fig:quantum_materials_pipeline}
\end{figure}
The SSF provides a more targeted route to the long-wavelength physics. It measures density--density correlations in momentum space and enters the two-body energy through Coulomb-induced density fluctuations. Since the Coulomb kernel strongly weights small momenta, the dominant residual two-body finite-size error is controlled by the infrared behavior of the SSF. This is why structure-factor corrections are so useful in QMC. In practice, however, extracting the low-momentum behavior from classical QMC can be delicate: the fermion sign problem restricts the accessible system sizes precisely in metallic and extreme-condition regimes, where finite-size effects are most important~\cite{Troyer_2005_sign_problem,Loh_1990_sign_problem,Foulkes_2001_QMC_solids,Assaad_2008_QMC_methods}.
\\

In our work we address this post-processing stage of fault-tolerant materials simulation: how to remove finite-size errors once a finite-cell ground-state energy has been obtained. The broader workflow still requires state preparation and ground-state energy estimation, and we assume access to a trial state with non-negligible overlap with the ground state. Our focus is the finite-size correction, which has so far largely remained outside the quantum workflow. Existing approaches repeat the full energy-estimation routine across a hierarchy of larger supercells or denser $k$-meshes~\cite{ivanov_2023_periodic,Rubin_bloch_2023,ivanov_2024_paw}. We replace this repeated use of the most expensive routine with a fault-tolerant computation of low-momentum samples of the SSF, from which the leading two-body finite-size correction can be reconstructed directly. The resulting strategy combines twist averaging for one-body shell effects with a quantum evaluation of the low-momentum SSF for the residual two-body correction. In this way, reaching the thermodynamic-limit is tied directly to the long-wavelength density correlations that control the finite-size bias, rather than to repeated ground-state energy calculations on larger simulation cells. This distinction is especially important for metallic and extreme-condition materials, where dense $k$-space resolution can otherwise amplify quantum resource estimates. Figure \ref{fig:quantum_materials_pipeline} gives a pictorial description of this procedure.
\\

Our work makes three main contributions:
\begin{enumerate}
    \item We provide a fault-tolerant procedure for estimating SSF using an amplified Hadamard test, together with adaptive global and local binary search protocols for extracting the infrared structure-factor data that enter the finite-size correction.

    \item We define the SSF operator, whose expectation value with respect to the ground state yields SSF. We express it in a Bloch-orbital basis, and construct its block encoding. This representation preserves crystal-momentum structure.

    \item We provide exact circuit architecture and the resulting resource estimates in the form of Toffoli and qubit count. We further show that the finite-size correction remains subleading relative to the main ground-state energy estimation routine. The correction has leading Toffoli scaling proportional to $(N_bN_k)^3$, but avoids the large plane-wave prefactor appearing in full Hamiltonian simulation~\cite{Rubin_bloch_2023,Bhardwaj:2026koe,ivanov_2024_paw}; its logical-qubit overhead scales only as $\mathcal{O}(N_bN_k)$.
\end{enumerate}
The paper is organized as follows. Section~\ref{sec: preliminaries} reviews finite-size mitigation in periodic simulations, emphasizing twist averaging, and introduces the static structure factor and its connection to long-wavelength density fluctuations and two-body corrections. Section~\ref{sec: methods} develops the binary-search-based infrared reconstruction procedures used to determine the two-body finite-size correction and presents the fault-tolerant Hadamard-test protocol for estimating $S(\mbf q)$. Section~\ref{sec: results} constructs $S(\mbf q)$ in the Bloch basis, gives its block encoding, and analyzes the cost of its computation. Finally, Section~\ref{sec: cost_comparison} compares the cost of the finite-size correction with that of the main ground-state energy estimation routine and summarizes the scaling advantage of treating finite-size mitigation as a quantum post-processing task.
\section{Preliminaries}
\label{sec: preliminaries}
\subsection{Twist averaging as the one-body component of finite-size error mitigation}
\label{subsec:finite_size_mitigation}

\begin{figure}[t]
\centering
\begin{tikzpicture}[scale=0.75,>=Latex]

\tikzset{
    bzcurrent/.style={draw=black!75, dashed, line width=1.0pt},
    bzold/.style={draw=gray!50, dashed, line width=0.9pt, opacity=0.30},
    ptcurrent/.style={circle, fill=black, inner sep=1.5pt},
    ptold/.style={circle, fill=gray!50, inner sep=1.4pt, opacity=0.30},
    sample/.style={circle, fill=blue!70, inner sep=1.8pt},
    band/.style={red!75, line width=1.0pt},
    arrow/.style={-{Stealth[length=1mm,width=2mm]}, line width=0.9pt},
    label/.style={font=\small},
    title/.style={font=\small\bfseries},
    box/.style={rounded corners=3pt, draw=gray!55, fill=gray!6}
}

\node[title] at (-5.1,2.85) {Twist $\theta_1$};

\begin{scope}[shift={(-5.1,0.6)}]
    \draw[bzcurrent] (90:1.35) \foreach \a in {150,210,270,330,30} { -- (\a:1.35)} -- cycle;

    \foreach \x/\y in {
        -0.78/0.42, -0.18/0.60, 0.48/0.50,
        -0.92/-0.05, -0.28/0.05, 0.32/-0.02, 0.90/0.12,
        -0.62/-0.56, -0.04/-0.65, 0.62/-0.50
    }{
        \node[ptcurrent] at (\x,\y) {};
    }

    \draw[band]
      (-1.05,-0.12)
      .. controls (-0.58,0.12) and (-0.08,-0.32) .. (0.36,-0.18)
      .. controls (0.72,-0.05) and (0.90,0.10) .. (1.05,0.28);

    \node[sample] at (-0.82,-0.02) {};
    \node[sample] at (-0.22,-0.07) {};
    \node[sample] at (0.76,0.04) {};
\end{scope}

\node[title] at (0,2.85) {Twist $\theta_2$};

\begin{scope}[shift={(0,0.6)}]
    \begin{scope}[shift={(-0.16,0.10)}]
        \draw[bzold] (90:1.35) \foreach \a in {150,210,270,330,30} { -- (\a:1.35)} -- cycle;
        \foreach \x/\y in {
            -0.78/0.42, -0.18/0.60, 0.48/0.50,
            -0.92/-0.05, -0.28/0.05, 0.32/-0.02, 0.90/0.12,
            -0.62/-0.56, -0.04/-0.65, 0.62/-0.50
        }{
            \node[ptold] at (\x,\y) {};
        }
    \end{scope}

    \draw[bzcurrent] (90:1.35) \foreach \a in {150,210,270,330,30} { -- (\a:1.35)} -- cycle;
    \foreach \x/\y in {
        -0.60/0.58, 0.00/0.46, 0.66/0.62,
        -0.76/0.10, -0.10/0.14, 0.50/0.08, 0.98/0.22,
        -0.42/-0.42, 0.18/-0.50, 0.76/-0.34
    }{
        \node[ptcurrent] at (\x,\y) {};
    }

    \draw[band]
      (-1.05,-0.12)
      .. controls (-0.58,0.12) and (-0.08,-0.32) .. (0.36,-0.18)
      .. controls (0.72,-0.05) and (0.90,0.10) .. (1.05,0.28);

    \node[sample] at (-0.58,0.04) {};
    \node[sample] at (0.12,-0.20) {};
    \node[sample] at (0.94,0.18) {};
\end{scope}

\node[title] at (5.1,2.85) {Twist $\theta_3$};

\begin{scope}[shift={(5.1,0.6)}]
    \begin{scope}[shift={(-0.28,0.16)}]
        \draw[bzold] (90:1.35) \foreach \a in {150,210,270,330,30} { -- (\a:1.35)} -- cycle;
        \foreach \x/\y in {
            -0.78/0.42, -0.18/0.60, 0.48/0.50,
            -0.92/-0.05, -0.28/0.05, 0.32/-0.02, 0.90/0.12,
            -0.62/-0.56, -0.04/-0.65, 0.62/-0.50
        }{
            \node[ptold] at (\x,\y) {};
        }
    \end{scope}

    \begin{scope}[shift={(-0.10,0.05)}]
        \draw[bzold] (90:1.35) \foreach \a in {150,210,270,330,30} { -- (\a:1.35)} -- cycle;
        \foreach \x/\y in {
            -0.60/0.58, 0.00/0.46, 0.66/0.62,
            -0.76/0.10, -0.10/0.14, 0.50/0.08, 0.98/0.22,
            -0.42/-0.42, 0.18/-0.50, 0.76/-0.34
        }{
            \node[ptold] at (\x,\y) {};
        }
    \end{scope}

    \draw[bzcurrent] (90:1.35) \foreach \a in {150,210,270,330,30} { -- (\a:1.35)} -- cycle;
    \foreach \x/\y in {
        -0.92/0.28, -0.34/0.50, 0.28/0.40,
        -1.00/-0.18, -0.40/-0.04, 0.20/-0.12, 0.80/0.02,
        -0.70/-0.72, -0.10/-0.58, 0.50/-0.66
    }{
        \node[ptcurrent] at (\x,\y) {};
    }

    \draw[band]
      (-1.05,-0.12)
      .. controls (-0.58,0.12) and (-0.08,-0.32) .. (0.36,-0.18)
      .. controls (0.72,-0.05) and (0.90,0.10) .. (1.05,0.28);

    \node[sample] at (-0.98,-0.08) {};
    \node[sample] at (-0.38,-0.12) {};
    \node[sample] at (0.58,-0.05) {};
\end{scope}

\node[label] at (0,-1.40) {Average over twists};
\draw[arrow] (0,-1.58) -- (0,-2.25);

\node[title] at (0,-2.95) {Twist-averaged sampling};

\begin{scope}[shift={(0,-5.0)}]
    \node[box, minimum width=8cm, minimum height=3.5cm] at (0,0) {};

    \begin{scope}[shift={(-0.25,0.48)}]
        \draw[bzold] (90:1.10) \foreach \a in {150,210,270,330,30} { -- (\a:1.10)} -- cycle;
    \end{scope}
    \begin{scope}[shift={(0.00,0.30)}]
        \draw[bzold] (90:1.10) \foreach \a in {150,210,270,330,30} { -- (\a:1.10)} -- cycle;
    \end{scope}
    \begin{scope}[shift={(0.25,0.12)}]
        \draw[bzold] (90:1.10) \foreach \a in {150,210,270,330,30} { -- (\a:1.10)} -- cycle;
    \end{scope}

    \begin{scope}[shift={(0,0.12)}]
        \foreach \x/\y in {
            -4.00/0.35, -3.55/0.05, -3.10/0.52, -2.70/0.12, -2.25/0.46,
            -1.85/0.02, -1.45/0.55, -1.05/0.18, -0.60/0.42, -0.10/0.10,
             0.35/0.48,  0.80/0.08,  1.20/0.44,  1.65/0.02,  2.10/0.50,
             2.55/0.12,  3.00/0.40,  3.45/0.02,  3.90/0.32,
            -3.70/-0.18, -3.15/-0.42, -2.55/-0.10, -2.00/-0.32, -1.35/-0.44,
            -0.75/-0.18, -0.10/-0.36,  0.55/-0.14,  1.10/-0.40,  1.75/-0.22,
             2.35/-0.44,  2.95/-0.16,  3.55/-0.34
        }{
            \node[sample] at (\x,\y) {};
        }

        \draw[band]
          (-4.55,0.00)
          .. controls (-3.40,0.32) and (-2.55,-0.18) .. (-1.35,0.05)
          .. controls (-0.25,0.24) and (0.80,0.46) .. (2.00,0.22)
          .. controls (2.90,0.06) and (3.75,-0.08) .. (4.45,0.18);
    \end{scope}

    \node[label, align=center] at (0,-1.5)
    {$\displaystyle
      \hat{E}(L,N_k)=\frac{1}{M}\sum_{\btheta} E_{\theta}(L,N_k)
      \;\approx\;
      \int_{\mathrm{BZ}} d^3k\, f(\mbf{k})$};
\end{scope}

\end{tikzpicture}
\caption{\small Schematic illustration of twist averaging. Each twist shifts the underlying finite $k$-point mesh relative to the previous one, leading to a different sampling of the same band-structure feature. Averaging the resulting energies over many such shifted meshes yields a finer sampling of the BZ, reducing the one-body shell effects, and improves the approximation to the continuum BZ average without enlarging the underlying simulation cell.\normalsize}
\label{fig:twist_averaging_tikz}
\end{figure}
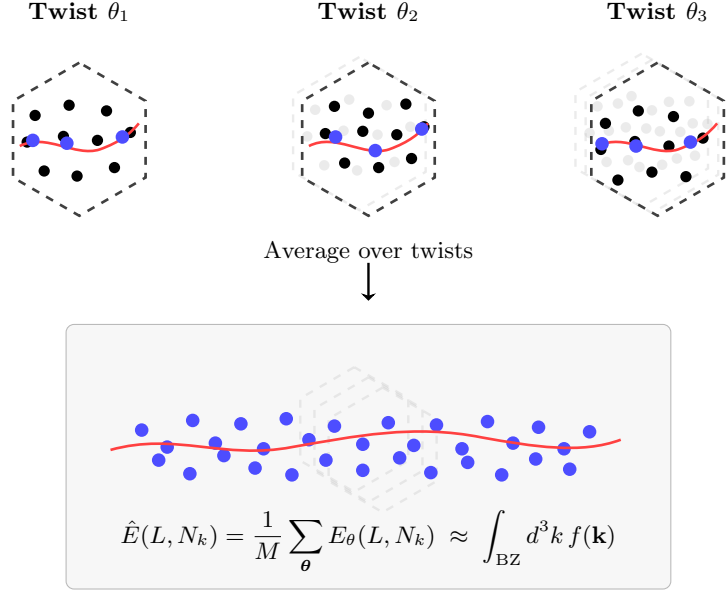

 For the present purposes, it is useful to distinguish between two broad sources of finite-size error. The first are one-body, or shell, effects arising from the discretization of the BZ. These are especially severe in metallic systems, where the energy depends sensitively on the occupation of states near the Fermi surface. The second are two-body finite-size effects associated with long-range Coulomb correlations. Once shell effects are sufficiently reduced, the remaining error is dominated by the long-wavelength physics of the interacting system and is naturally encoded in the small-$q$ behavior of the SSF $S(\mbf q)$.
To suppress the one-body contribution more economically, many-body methods such as QMC and full configuration interaction (FCI) commonly employ twist averaging \cite{Lin_2001,Shepherd_2014,Liao_2016,Shi_2021}. Rather than increasing the underlying simulation cell, one evaluates the many-body energy under twisted boundary conditions and averages over a set of twist vectors $\btheta$:
\begin{equation}
    \hat{E}(L,N_k)
    =
    \frac{1}{M}
    \sum_{\btheta}
    E_{\btheta}(L,N_k)~,
\end{equation}
where $M$ is the number of twists. If $\mathfrak{M}$ denotes a $\Gamma$-centered mesh with $N_k$ $k$-points, then the mesh associated with a twist $\btheta$ is
\begin{equation}
    \mathfrak{M}_{\theta}
    =
    \left\{
    \mbf{k}_{\theta}
    \;\middle|\;
    \mbf{k}_{\theta}
    =
    \mbf{k}
    +
    \frac{2\pi}{L}\btheta,
    \;
    \mbf{k}\in \mathfrak{M}
    \right\}~.
\end{equation}
where $L$ is the size of the Born-von Kármán (BvK) cell that describes the periodic solid.
Operationally, twist averaging samples the BZ more uniformly and thereby reduces the sensitivity of the energy to any single discretization of momentum space. In the one-body sector, this may be viewed as replacing the continuum integral
\begin{equation*}
    E_{\rm 1b}
    =
    \int_{{\rm BZ}} d^3k\, f(\mbf{k})
\end{equation*}
by an average over shifted quadratures,
\begin{equation*}
E_{\rm 1b} ~~\xmapsto{\mathrm{discretization}}~~
\frac{1}{\Omega\;M} \sum_{\btheta}\sum_{\mbf{k}\in \mathfrak{M}_{\theta}} f(\mbf{k})~,
\end{equation*}
where $\Omega$ is the volume of the BvK cell. This is particularly advantageous for metallic systems, where fine $k$-space resolution is otherwise needed near the Fermi surface. In practice, twist averaging can often reproduce the accuracy of a much denser mesh while retaining a much smaller underlying simulation cell. For example, in Ref. \cite{Azadi_2019}, metallic H, Li, and Al needed a very dense mean-field $k$-mesh to resolve the Fermi surface accurately: in particular a $24\times24\times24$ DFT mesh was used to determine the Fermi energy. By contrast, in Ref. \cite{Annaberdiyev_2024}, diffusion Monte Carlo studies of fcc Al with twist-averaged observables were found to converge for a $2\times 2\times 2$ $k$-mesh. The number of twists required to reach target accuracy $\epsilon$ depends on the twist-to-twist energy variation $\sigma_{\theta}$,
\begin{equation}
    M
    \sim
    \mathcal{O}\!\left(\frac{\sigma_{\theta}}{\epsilon}\right)^2,
    \qquad
    \sigma_{\theta}\equiv |E_{\theta}-E_{\theta'}|~,
\end{equation}
and this variation typically decreases with mesh size, often approximately as
\begin{equation}
    \sigma_{\theta}\sim \mathcal{O}(N_k^{-\mathfrak{a}})~,
\end{equation}
where the exponent $\mathfrak{a}$ is material dependent and can be especially favorable in insulators or weakly metallic systems \cite{Lin_2001, Azadi_2019, Annaberdiyev_2024}.
\\
From the standpoint of quantum resource optimisation, the key advantage is that twist averaging does not enlarge the system size itself. Each twist corresponds to an independent quantum calculation on the same underlying simulation cell, and the final average is performed entirely as a post-processing step. The logical qubit requirement is set by the underlying simulation cell and is therefore unaffected by the number of twist samples $M$. By contrast, the Toffoli cost increases sublinearly with $M$. This is because, to achieve an overall error $\epsilon$ in the twist-averaged energy, each individual twist energy need only be estimated to precision $\sqrt{M}\,\epsilon$. Since the cost of a single energy estimate typically scales as the inverse target precision, the total cost of the twist average scales as $\mathcal{O}(\sqrt{M}/\epsilon)$ rather than $\mathcal{O}(M/\epsilon)$. In this sense, twist averaging leaves the logical qubit count unchanged while increasing the gate cost only as the square root of the number of twist samples. This is fundamentally different from brute-force finite-size convergence, in which improving accuracy requires more $k$-points and therefore increases both the gate complexity and the qubit count of each energy-evaluation instance. A schematic illustration of the twist-averaging procedure is shown in Fig.~\ref{fig:twist_averaging_tikz}.

\vspace{5mm}
Once one-body component of the finite-size error is reduced via the above twist averaging procedure, the dominant contribution to the error is a two-body effect originating from long-range Coulomb correlations. These residual errors are concentrated in the infrared sector of momentum space and are controlled by the SSF. Twist averaging and structure-factor corrections should therefore be viewed as complementary rather than competing strategies: the former suppresses one-body discretization errors, while the latter reconstructs the leading two-body finite-size correction.
\\
This separation of roles motivates the overall strategy developed in our work. We first compute the twist-averaged finite-system energy $\hat{E}(L,N_k)$ in order to suppress one-body shell effects without enlarging the problem size for our quantum simulation instance. We then estimate the SSF for a small set of low-momentum vectors $\mbf q$ below an infrared cutoff $q_t$, and use these data to reconstruct the residual two-body finite-size correction $\Delta V(L,N_k)$. The resulting thermodynamic-limit energy density is estimated as
\begin{equation}
    E_{\infty}
    =
    \frac{1}{\Omega}\hat{E}(L,N_k)
    +
    \Delta V(L,N_k)~.
\end{equation}
In what follows we absorb the BvK simulation cell volume factor $\Omega$ into the target chemical accuracy $\epsilon$. In this way, the thermodynamic-limit problem is decomposed into two computationally distinct tasks: a twist-averaged energy calculation that controls one-body shell effects, and an infrared reconstruction of $\Delta V$ from a small number of low-momentum values of $S(\mbf q)$ that captures the leading two-body contribution. As we will show in the following sections, as compared with down-sampling over increasingly large supercells or denser $k$-meshes, this hybrid strategy is much better aligned with fault-tolerant resource constraints. 
\subsection{Introducing the static structure factor}
\label{subsec:static_structure_factor}

 To isolate the long-wavelength correlations relevant for finite-size effects, we begin from the many-body Hamiltonian,
\begin{equation}
    \hat{H} = \hat{T} + \hat{U} + \hat{V},
\end{equation}
where $\hat{T}$, $\hat{U}$, and $\hat{V}$ denote the kinetic, one-body, and two-body contributions, respectively. We keep these terms at a fairly general level rather than specializing immediately to the conventional electronic Hamiltonian. As a result, the discussion up to the end of Section~\ref{sec: quantum_algorithm} applies to a broader class of many-body Hamiltonians sharing this structure. Nonetheless, in the field-operator representation these terms in atomic units are
\begin{align}
    \hat{T}
    &=
    -\frac{1}{2}\int_{\Omega} d^3\mbf{r}\;
    \hat{\psi}^{\dg}(\mbf{r})\nabla^2 \hat{\psi}(\mbf{r})~, \\
    \hat{U}
    &=
    \int_{\Omega} d^3\mbf{r}\;
    \hat{\psi}^{\dg}(\mbf{r}) U(\mbf{r}) \hat{\psi}(\mbf{r})~, \\
    \hat{V}
    &=
    \frac{1}{2}\int_{\Omega\times\Omega} d^3\mbf{r}\, d^3\mbf{r}'\;
    \hat{\psi}^{\dg}(\mbf{r})\hat{\psi}(\mbf{r})
    V(\mbf{r}'-\mbf{r})
    \hat{\psi}^{\dg}(\mbf{r}')\hat{\psi}(\mbf{r}')~.
\end{align}
To isolate the long-wavelength physics, it is natural to work in Fourier space and introduce the operators
\begin{align}
    \hat{\rho}_{\mbf{q}} &= \int_{\Omega}d^3\mbf{r}~e^{i\mbf{q}\cdot\mbf{r}}\hat{\psi}^{\dg}(\mbf{r})\hat{\psi}(\mbf{r}), \label{eq: rho_matrix}\\
    \hat{\gamma}_{\mbf{q}} &= \int_{\Omega\times \Omega}d^3\mbf{r}~d^3\mbf{r}'~e^{i\mbf{q}\cdot(\mbf{r}-\mbf{r}')}\hat{\psi}^{\dg}(\mbf{r})\hat{\psi}(\mbf{r}') \label{eq: gamma_matrix}.
\end{align}
Here $\hat{\rho}_{\mbf q}$ is the \emph{density operator} in the reciprocal space, while $\hat{\gamma}_{\mbf q}$ is the corresponding one-body quantity entering the kinetic term. In terms of these operators, the Hamiltonian takes the compact form
\begin{equation*}
    \hat{H} = \frac{1}{2}\int_{\tilde{\Omega}}\frac{d^3\mbf{q}}{(2\pi)^3}~\mbf{q}^2\hat{\gamma}_{\mbf{q}}
    ~+~\int_{\tilde{\Omega}}\frac{d^3\mbf{q}}{(2\pi)^3}~u(\mbf{q})\hat{\rho}_{\mbf{q}}
    ~+~\frac{1}{2}\int_{\tilde{\Omega}}\frac{d^3\mbf{q}}{(2\pi)^3}~v(\mbf{q})\hat{\rho}^{\dg}_{\mbf{q}}\hat{\rho}_{\mbf{q}}~,
\end{equation*}
where $u(\mbf{q})$ and $v(\mbf{q})$ are the Fourier transforms of the one-body and two-body potentials respectively.
This form makes clear that the two-body interaction is governed by density fluctuations through $\hat{\rho}^{\dg}_{\mbf q}\hat{\rho}_{\mbf q}$. The corresponding SSF is defined as
\begin{equation}
    S(\mbf{q})=\frac{1}{\Omega}\langle \psi_0 \,|\, \hat S_{\mbf q}\,|\,\psi_0\rangle~.
\end{equation}
where $\hat S_{\mbf k} = \hat{\rho}^{\dg}_{\mbf{k}}\hat{\rho}_{\mbf{k}}$ and $|\psi_0\rangle$ is the ground state of the system.
It is the standard momentum-space measure of density--density correlations \cite{Dobard_2013,Pereira_2007,Shukla_1999}. Its role becomes explicit when the ground-state energy density is written:
\begin{equation}
    E_{\infty} = \frac{1}{2}\int_{\tilde{\Omega}}\frac{d^3\mbf{q}}{(2\pi)^3}~\mbf{q}^2\gamma(\mbf{q})
    ~+~\int_{\tilde{\Omega}}\frac{d^3\mbf{q}}{(2\pi)^3}~u(\mbf{q})\rho(\mbf{q})
    ~+~\frac{1}{2}\int_{\tilde{\Omega}}\frac{d^3\mbf{q}}{(2\pi)^3}~v(\mbf{q})\left(S(\mbf{q})-1\right)~,
\end{equation}
where $\gamma(\mbf{q})$ and $\rho(\mbf{q})$ are the ground state expectation values of the corresponding operators defined in Eq. \eqref{eq: gamma_matrix} and \eqref{eq: rho_matrix} respectively. The final term is the one relevant for the present work: it shows that the interaction energy is obtained by weighting the structure factor (which we have also subtracted by unity to account for the exclusion of uncorrelated effects) by the interaction kernel $v(\mbf q)$. For Coulomb systems, this weighting is strongest at small $|\mbf q|$, so the long-wavelength behavior of $S(\mbf q)$ controls the leading residual two-body finite-size error. At a physical level, $S(\mbf q)$ measures density correlations at wavelength $2\pi/|\mbf q|$. With the present $1/\Omega$ normalization, an uncorrelated system satisfies
$S(\mbf q)=n$ for $\mbf q\neq 0$, where $n$ is the electron number density. This corresponds to the usual dimensionless static structure factor being equal to unity, as expected for Poissonian density fluctuations. Correlations modify this behavior by suppressing or enhancing fluctuations depending on the relevant length scale. In particular, the small-$q$ limit probes collective, long-range density response and is therefore the regime that connects a finite periodic simulation cell to the thermodynamic limit. It is worth noting that related expressions in the QMC literature are often written as sums over reciprocal-lattice vectors rather than as integrals over continuum Fourier modes. This is because QMC commonly replaces the bare Coulomb interaction by a periodized interaction, such as the Ewald form \cite{Drummond_2008}, whose Fourier representation is naturally supported on the reciprocal lattice of the finite cell. In the present work we instead retain the bare Coulomb kernel and impose periodicity through the Bloch-orbital basis. The bookkeeping is therefore different, but the physical conclusion is the same: the leading two-body finite-size correction is controlled by the infrared structure of $S(\mbf q)$. A detailed discussion of the equivalence between the QMC formulation and the present approach to finite-size corrections is provided in Appendix~\ref{app: QMC_Bloch_equivalence} in the context of homogenous electron gas.
\subsection{Finite size corrections from the static structure factor}
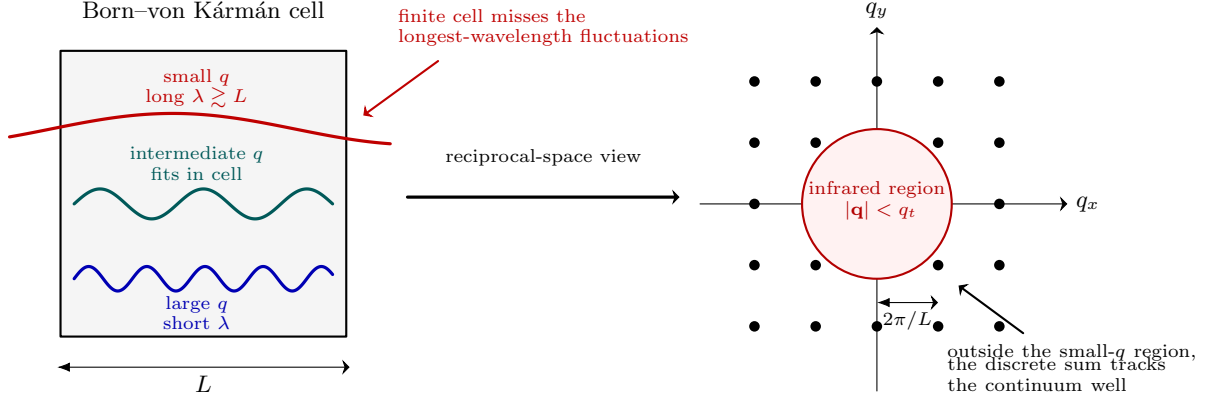
\begin{figure}[ht!]
    \centering
    \begin{tikzpicture}[
        scale=0.9,
        line cap=round,
        line join=round,
        >=Latex,
        font=\small
    ]
    
    \tikzset{
        cellface/.style={fill=gray!8, draw=black, thick},
        shortmode/.style={blue!70!black, very thick},
        mediummode/.style={teal!70!black, very thick},
        longmode/.style={red!75!black, very thick},
        qpoint/.style={circle, fill=black, inner sep=1.4pt}
    }
    
    \coordinate (A) at (0,0);
    \coordinate (B) at (4.2,0);
    \coordinate (C) at (4.2,4.2);
    \coordinate (D) at (0,4.2);
    
    \fill[gray!10] (A)--(B)--(C)--(D)--cycle;
    \draw[cellface] (A)--(B)--(C)--(D)--cycle;
    
    \node[above] at ($(D)!0.5!(C)+(0,0.35)$) {Born--von K\'arm\'an cell};
    \draw[<-{Stealth[length=1mm,width=2mm]}] (-0.05,-0.45) -- node[below] {$L$} (4.25,-0.45);
    
    \begin{scope}
        \clip (A) rectangle (C);
    
        \draw[shortmode]
        plot[samples=220, domain=0.20:4.00]
        (\x,{0.85 + 0.18*sin(1600*(\x-0.20)/3.80)});
        \node[blue!70!black, align=center] at (2.0,0.35)
        {\scriptsize large $q$\\[-1mm]\scriptsize short $\lambda$};
    
        \draw[mediummode]
        plot[samples=220, domain=0.20:4.00]
        (\x,{1.95 + 0.22*sin(900*(\x-0.20)/3.80)});
        \node[teal!70!black, align=center] at (2.0,2.55)
        {\scriptsize intermediate $q$\\[-1mm]\scriptsize fits in cell};
    \end{scope}
    
    \draw[longmode]
    plot[samples=250, domain=-0.75:4.85]
    (\x,{3.05 + 0.23*sin(280*(\x+0.15)/5.60)});
    \node[red!75!black, align=center] at (2.0,3.65)
    {\scriptsize small $q$\\[-1mm]\scriptsize long $\lambda \gtrsim L$};
    
    \draw[-{Stealth[length=1mm,width=2mm]}, thick, red!75!black] (5.55,4.05) -- (4.45,3.25);
    \node[red!75!black, align=left] at (7.1,4.55)
    {\scriptsize finite cell misses the\\[-1mm]
     \scriptsize longest-wavelength fluctuations};
    
    \draw[-{Stealth[length=1mm,width=2mm]}, very thick] (5.1,2.05) -- (9.1,2.05);
    \node[above] at (7.1,2.38) {\scriptsize reciprocal-space view};
    
    \coordinate (O) at (12.0,1.95);
    
    \draw[-{Stealth[length=1mm,width=2mm]}] (9.4,1.95) -- (14.8,1.95) node[right] {$q_x$};
    \draw[-{Stealth[length=1mm,width=2mm]}] (12.0,-0.8) -- (12.0,4.55) node[above] {$q_y$};
    
    \foreach \i in {-2,-1,0,1,2}
    {
        \foreach \j in {-2,-1,0,1,2}
        {
            \node[qpoint] at ($(O)+(\i*0.90,\j*0.90)$) {};
        }
    }
    
    \fill[red!5] (O) circle (1.1);
    \draw[red!70!black, thick] (O) circle (1.1);
    
    \node[red!70!black, align=left] at (12.0,2.00)
    {\scriptsize infrared region\\[-1mm]
     \scriptsize ~~~~~$|\mbf q|<q_t$};
    
    \draw[<-{Stealth[length=1mm,width=2mm]}] ($(O)+(0,-1.45)$) -- node[below] {\scriptsize $2\pi/L$} ($(O)+(0.90,-1.45)$);
    \draw[-{Stealth[length=1mm,width=2mm]}, thick, black] (14.15,0.05) -- ($(O)-(-1.20,1.20)$);
    \node[align=left, text width=3.8cm] at (15.15,-0.45)
    {\scriptsize outside the small-$q$ region,\\[-1mm]
     \scriptsize the discrete sum tracks\\[-1mm]
     \scriptsize the continuum well};
    
    \end{tikzpicture}
    \caption{\small Schematic illustration of why the leading two-body finite-size correction is dominated by the small-$q$ sector of the SSF. In a finite Born--von K\'arm\'an cell of linear size $L$, short- and intermediate-wavelength fluctuations are well represented, whereas the longest-wavelength modes are only incompletely resolved. In reciprocal space, this appears as a mismatch between the continuum interaction energy and its discrete $k$-mesh representation primarily in the infrared region $|\mbf q|<q_t$, while the discrete sum already tracks the continuum well at larger $|\mbf q|$.\normalsize}
    \label{fig:deltaV_smallq}
\end{figure}

We now consider discretizing the real- and reciprocal-space. Throughout our work, all real- and reciprocal-space quantities are defined with respect to the periodic simulation supercell. We consider a cubic $N_a$-atom supercell together with a cubic $(N_k^{1/3}\times N_k^{1/3}\times N_k^{1/3})$ sampling of the BZ, so that the total BvK cell contains $N_aN_k$ primitive repeats and has volume $\Omega = N_aN_k\Omega_{\rm unit}$. 
\\
The role of the SSF in the finite-size problem follows directly from the reciprocal-space form of the energy. As discussed in the previous subsection, the two-body interaction energy is obtained by weighting the density--density correlations $S(\mbf q)$ by the interaction kernel $v(\mbf q)$. In a finite periodic cell, however, the continuum of allowed momenta is replaced by a discrete mesh $\mathfrak{M}$. The resulting finite-size error in the interaction energy is therefore the difference between the continuum expression and its discrete representation,
\begin{align}
    \Delta V
    &= \frac{1}{2}\int_{\rm BZ}\frac{d^3\mbf{q}}{(2\pi)^3}~v(\mbf{q})\left(S(\mbf{q})-1\right)~-~\frac{1}{2\Omega~M}\sum_{\btheta}\sum_{\substack{\mbf{q}+\mbf{G}\in \mathfrak{M};\\ |\mbf{q}+\mbf{G}|\leq|\mbf{q}_{\rm cut}|}}v(\mbf{q}+\mbf{G})\left(S^{(\btheta)}_{N_k}(\mbf{q}+\mbf{G})-1\right)~,\label{eq:delta_V_full}
\end{align}
where we now introduce the discrete SSF for given twist $\btheta$ as follows,
\begin{equation}
    S^{(\btheta)}_{N_k}(\mbf{q})=\frac{1}{\Omega}\langle \psi_0^{(\btheta)} \,|\, \hat S^{(\btheta)}_{\mbf q}\,|\,\psi_0^{(\btheta)}\rangle~,\label{eq: SSF_def_twisted}
\end{equation}
where $\psi^{(\btheta)}_0$ is the ground state with respect to the twist $\btheta$ and similarly $\hat S^{(\btheta)} = (\hat \rho^{(\btheta)})^{\dg}\hat \rho^{(\btheta)}$. Furthermore we have the added subscript serving to distinguish it from the continuum version $S(\mbf{q})$. It is worth noting that, as in the QMC literature \cite{Chiesa2007RPA_FSE,Dornheim_2021,Drummond_2008}, the discritized sum still has the reciprocal-lattice sums rather than a simple $k$-mesh sum. This is because of the periodized interactions in the reciprocal space. The expression above makes clear that the correction is controlled by the mismatch between the continuum description and the discrete momentum grid. For Coulomb systems, this mismatch is dominated by the infrared sector. The large-$|\mbf q|$ region is typically already well described by the discrete sum, whereas the leading error comes from the longest-wavelength fluctuations, corresponding to the smallest nonzero momenta. It is therefore natural to restrict the correction to a low-momentum window $|\mbf q|\le q_t$, where the continuum and discrete descriptions differ most:
\begin{equation}
    \Delta V
    \approx
    \frac{1}{2}\int_{|\mbf{q}|\le q_t}\frac{d^3\mbf{q}}{(2\pi)^3}\,v(\mbf{q})\left(S(\mbf{q})-1\right)
    -\frac{1}{2\Omega\;M}\sum_{\btheta}\sum_{\substack{\mbf{q}\in \mathfrak{M};\\ |\mbf{q}|\leq q_t}}v(\mbf{q})\left(S^{(\btheta)}_{N_k}(\mbf{q})-1\right)~,
    \label{eq:delta_V_ir}
\end{equation}
where for long-wavelength finite-size corrections one typically needs only the smallest $|\mbf{q}|$ which implies $\mbf{G} = 0$ and $\mbf{q}$ lies in some small sub-region of BZ. Eq.~\eqref{eq:delta_V_ir} is the central observation behind the present approach: the leading two-body finite-size correction is determined primarily by the small-$q$ behavior of the SSF. In practice, one therefore does not need the full momentum dependence of $S(\mbf q)$, but only its behavior in the infrared region. Fig.~\ref{fig:deltaV_smallq} illustrates this point schematically. This perspective also clarifies why the SSF is the natural object for finite-size corrections. Since $S(\mbf q)$ measures density fluctuations at wavelength $2\pi/|\mbf q|$, its small-$q$ limit probes the collective, long-range correlations that are most strongly distorted by a finite simulation cell. The leading residual finite-size error in the interaction energy is therefore a direct consequence of the fact that these longest-wavelength modes are either missing or poorly resolved on a finite momentum grid. 
\\

By contrast, the finite-size error in the one-body contribution $\Delta U$ is primarily a residual BZ quadrature error. The potentially dominant long-wavelength contribution cancels by charge neutrality, so $\Delta U$ does not receive the same infrared enhancement. As a result, once the $k$-mesh is reasonably dense, the one-body correction is sub-leading, whereas the two-body term $\Delta V$ remains the dominant source of finite-size error. A detailed analysis supporting this conclusion is provided in Appendix~\ref{app: DeltaU_is_subleading}.

\section{Methods}
\label{sec: methods}
\subsection{Adaptive extraction of the infrared contribution}
\label{sec: infrared_behavior_cutoff_procedure}

The finite-size correction developed in our work depends on the long-wavelength behavior of the SSF. The central practical task is therefore to identify a momentum window in which the measured $S^{(\btheta)}_{N_k}(\mbf q)$ has reached its asymptotic infrared form. The infrared extraction step is central to the overall resource estimate, because it determines how many fault-tolerant evaluations of the SSF are required. In this subsection, we introduce a practical strategy for reconstructing the low-\(q\) behavior of \(S(\mbf q)\) from a limited set of sampled momentum shells. We first state the infrared assumption, briefly place it in the context of established QMC practice, and then introduce two adaptive variants of the cutoff-selection procedure: a global search and a local windowed search. In the infrared, the SSF is expected to follow a power law, 
\begin{equation}
    S(\mbf{q}) \sim \alpha q^{\beta},
    \qquad q:=|\mbf q|~,
    \label{eq: IR_assumption}
\end{equation}
with $\beta=1$ for independent-particle or Hartree--Fock metals and $\beta=2$ for insulators. We note, however, that for the fully interacting charge SSF of a 3D Coulomb metal, long-range Coulomb screening leads to $\beta = 2$, so the distinction from an insulator appears mainly in the coefficient $\alpha$ rather than in $\beta$ \cite{PhysRevLett.97.076404, Drummond_2008, PhysRevB.85.125125}. For more details and references on the validity of our IR ansatz in Eq. \eqref{eq: IR_assumption} see Appendix \ref{app: note on the IR ansatz}. This infrared ansatz becomes reliable once $N_k$ is large enough that the subleading term falls to the level of the error budget assigned to $S(\mbf q)$, i.e. $\sigma_S \approx \gamma q^{\beta+1},$
where $\gamma$ is the coefficient of the first subleading correction. Equivalently, one requires
\begin{equation}
    N_k \gtrsim \frac{(2\pi)^3}{\Omega_{\rm sc}}
    \left(\frac{\gamma}{\sigma_S}\right)^{\frac{3}{\beta+1}}~\label{eq: N_k_constraint}.
\end{equation}
where $\Omega_{\rm sc} = N_a\Omega_{\rm unit}$ is the supercell volume.  In this regime, the finite-size correction $\Delta V$ is determined primarily by the leading infrared coefficient $\alpha$. In practice, however, one must still determine the largest momentum cutoff $q_t$ for which the sampled data remain in the asymptotic regime: if $q_t$ is chosen too small, the fit is statistically unstable because too few modes are included; if it is chosen too large, non-universal finite-$q$ structure contaminates the infrared fit.
\\

As previously mentioned, a common strategy in the QMC literature is to use RPA as a model for the low-$q$ response and thereby infer the missing contribution to the finite-size correction \cite{PhysRevLett.97.076404,Chiesa2007RPA_FSE,Dornheim_2021}. This is often effective when long-range screening dominates, but it can become unreliable when short-range strong correlations are important. For this reason, we adopt a complementary model-independent procedure that determines $q_t$ directly from the measured behavior of $S^{(\btheta)}_{N_k}(\mbf q)$.
\\

For fixed value of the twist $\btheta$, we introduce two variants of this cut-off selection procedure, the \emph{global adaptive search} and the \emph{local windowed search}. Both variants of the procedure use the same basic stability criterion. For a candidate cutoff indexed by $m$, let $\alpha(m)$ denote the slope extracted from the corresponding low-$q$ fit and $\sigma(m)$ its uncertainty. We regard two candidate cutoffs $m$ and $m'$ as statistically consistent if
\begin{equation}
    \big|\alpha(m)-\alpha(m')\big|
    \le
    z\sqrt{\sigma(m)^2+\sigma(m')^2-2\text{Cov}(m,m')}~,
    \label{eq:valid_predicate}
\end{equation}
where $z$ is a chosen confidence parameter. Eq.~\eqref{eq:valid_predicate} tests whether enlarging the fitting window changes the inferred infrared slope by more than allowed by its statistical uncertainty. For convenience, Table~\ref{tab:cutoff_search_summary} summarizes the main features and practical trade-offs of the two search strategies before we describe them in detail.
\begin{table*}[t]
\centering
\begin{tcolorbox}[
    colback=black!4,
    colframe=black!35,
    boxrule=0.5pt,
    arc=3mm,
    left=4mm,
    right=4mm,
    top=2mm,
    bottom=2mm,
    width=0.96\textwidth
]
\begin{threeparttable}
\caption{Comparison of the two adaptive procedures used to identify the infrared fitting window. Both methods return a stable cutoff $q_t^*$ and the associated slope $\alpha^*$, but differ in how the low-$q$ data are used and in the number of required evaluations of $S^{(\btheta)}_{N_k}(\mbf q)$.}
\label{tab:cutoff_search_summary}
\small
\renewcommand{\arraystretch}{1.2}

\setlength{\tabcolsep}{3pt}
\arrayrulecolor{black!35}
\newcommand{\cutoffcell}[2]{\parbox[t]{#1}{\raggedright #2}}
\begin{tabular}{|l|l|l|l|l|}
\hline
\cutoffcell{2.1cm}{Method} &
\cutoffcell{2.7cm}{Fit window} &
\cutoffcell{2.4cm}{$S^{(\btheta)}_{N_k}$ evaluations} &
\cutoffcell{3.0cm}{Strength} &
\cutoffcell{3.0cm}{Limitation} \\
\hline
\cutoffcell{2.1cm}{Global adaptive search} &
\cutoffcell{2.7cm}{All sampled shells with $q \le q_t$} &
\cutoffcell{2.4cm}{Worst case: $\mathcal{O}(N_k^{\frac{2}{3}})$} &
\cutoffcell{3.0cm}{Most direct and robust identification of the infrared regime} &
\cutoffcell{3.0cm}{Computationally more expensive, since it requires a search over the full low-$q$ region} \\
\hline
\cutoffcell{2.1cm}{Local windowed search} &
\cutoffcell{2.7cm}{Sliding window of $w$ neighboring low-$q$ shells} &
\cutoffcell{2.4cm}{Worst case: $\mathcal{O}(\frac{2}{3}w\log N_k)$} &
\cutoffcell{3.0cm}{Reduces the number of new shell evaluations by reusing previously computed data} &
\cutoffcell{3.0cm}{Requires additional classical storage and bookkeeping} \\
\hline
\end{tabular}
\end{threeparttable}
\end{tcolorbox}
\end{table*}

\paragraph{Shared setup.}
Let $q_{\min}=2\pi/L$ denote the smallest nonzero momentum accessible in a cubic supercell of linear size $L$, and define candidate infrared cutoffs by $q_t(m)=m q_{\min}$, where $m$ labels momentum shells. For each candidate cutoff, we extract an infrared slope $\alpha(m)$ from the measured low-$q$ data and assess its stability using Eq.~\eqref{eq:valid_predicate}. The goal is to identify the largest cutoff $q_t^*=q_t(m^*)$ for which the inferred slope remains statistically stable, and to use the corresponding $\alpha^*=\alpha(m^*)$ in the finite-size correction.
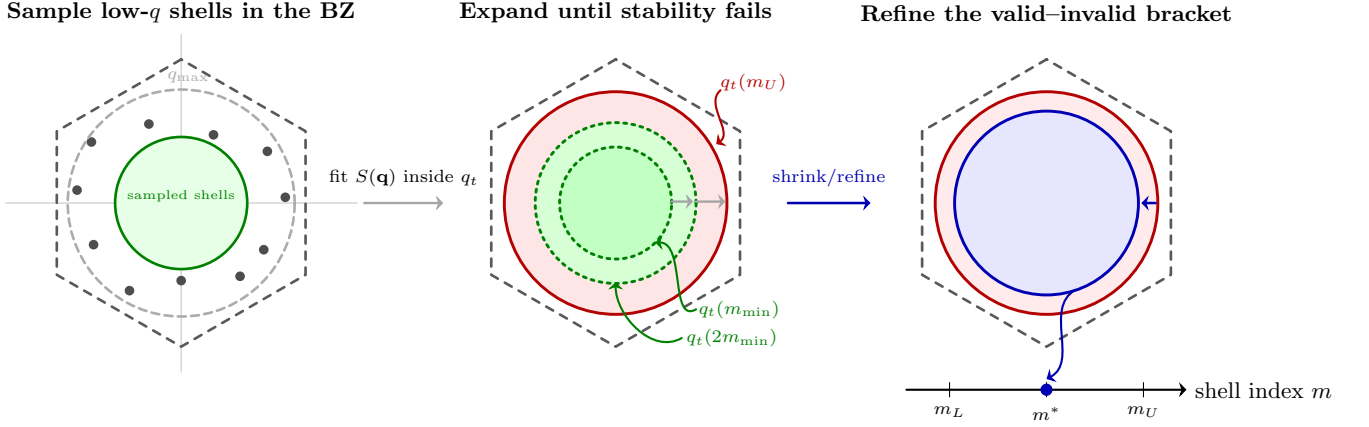
\begin{figure}[htbp]
\centering
\begin{tikzpicture}[
    scale=0.95,
    transform shape,
    line cap=round,
    line join=round,
    >=Latex,
    font=\small
]

\tikzset{
    bz/.style={draw=black!65, dashed, line width=1.0pt},
    meshpt/.style={circle, fill=black!70, inner sep=1.35pt},
    qmaxball/.style={draw=gray!65, densely dashed, line width=1.0pt},
    validball/.style={draw=green!50!black, fill=green!10, line width=1.0pt},
    validoutline/.style={draw=green!50!black, line width=1.0pt},
    invalidball/.style={draw=red!70!black, fill=red!8, line width=1.1pt},
    optball/.style={draw=blue!70!black, fill=blue!10, line width=1.1pt},
    panelarrow/.style={-{Stealth[length=1mm,width=2mm]}, line width=0.9pt, gray!70},
    refinearrow/.style={-{Stealth[length=1mm,width=2mm]}, line width=0.95pt, blue!70!black},
    lab/.style={font=\small\bfseries},
    tinylabel/.style={font=\scriptsize}
}

\node[lab] at (2.1,4.9) {Sample low-$q$ shells in the BZ};
\node[lab] at (8.15,4.9) {Expand until stability fails};
\node[lab] at (14.15,4.9) {Refine the valid--invalid bracket};

\begin{scope}[shift={(2.1,2.25)}]
    \draw[gray!45, thin] (-2.45,0) -- (2.45,0);
    \draw[gray!45, thin] (0,-2.35) -- (0,2.35);

    \draw[bz] (90:2.0) \foreach \a in {150,210,270,330,30} { -- (\a:2.0)} -- cycle;

    \foreach \x/\y in {
    -1.25/0.85, -0.45/1.10, 0.45/0.95, 1.20/0.72,
    -1.45/0.18, -0.70/0.22, 0.00/0.25, 0.72/0.18, 1.45/0.08,
    -1.22/-0.58, -0.42/-0.48, 0.50/-0.56, 1.15/-0.65,
    -0.72/-1.22, 0.00/-1.08, 0.82/-1.02
    }{
        \node[meshpt] at (\x,\y) {};
    }

    \draw[qmaxball] (0,0) circle (1.58);
    \node[tinylabel, gray!65] at (0.10,1.76) {$q_{\max}$};

    \draw[validball] (0,0) circle (0.92);
    \node[tinylabel, green!50!black] at (0.00,0.10) {\tiny sampled shells};
\end{scope}

\draw[panelarrow] (4.65,2.25) -- (5.75,2.25);
\node[tinylabel] at (5.20,2.60) {fit $S(\mbf q)$ inside $q_t$};

\begin{scope}[shift={(8.15,2.25)}]
    \draw[bz] (90:2.0) \foreach \a in {150,210,270,330,30} { -- (\a:2.0)} -- cycle;

    \fill[red!10]   (0,0) circle (1.55);
    \fill[green!16] (0,0) circle (1.12);
    \fill[green!24] (0,0) circle (0.78);

    \draw[green!50!black, dotted, line width=1.1pt] (0,0) circle (0.78);
    \draw[green!50!black, dotted, line width=1.1pt] (0,0) circle (1.12);
    \draw[red!70!black, line width=1.1pt]           (0,0) circle (1.55);

    \draw[panelarrow] (0.78,0.02) -- (1.08,0.02);
    \draw[panelarrow] (1.12,0.02) -- (1.53,0.02);

    \node[tinylabel, green!50!black] at (1.72,-1.48) {$q_t(m_{\min})$};
    \draw[-{Stealth[length=1mm,width=2mm]}, green!50!black, line width=0.8pt]
        (1.12,-1.48) to[out=210,in=0] (0.55,-0.55);

    \node[tinylabel, green!50!black] at (1.62,-1.88) {$q_t(2m_{\min})$};
    \draw[-{Stealth[length=1mm,width=2mm]}, green!50!black, line width=0.8pt]
        (0.92,-1.88) to[out=210,in=-90] (0, -1.10);

    \node[tinylabel, red!70!black] at (1.92,1.68) {$q_t(m_U)$};
    \draw[-{Stealth[length=1mm,width=2mm]}, red!70!black, line width=0.8pt]
        (1.46,1.57) to[out=220,in=35] (1.40,0.82);
\end{scope}

\draw[refinearrow] (10.55,2.25) -- (11.70,2.25);
\node[tinylabel, blue!70!black] at (11.12,2.60) {shrink/refine};

\begin{scope}[shift={(14.15,2.25)}]
    \draw[bz] (90:2.0) \foreach \a in {150,210,270,330,30} { -- (\a:2.0)} -- cycle;

    \draw[green!50!black, dotted, line width=1.1pt] (0,0) circle (1.02);
    \fill[red!8]   (0,0) circle (1.55);
    \draw[red!70!black, line width=1.1pt]           (0,0) circle (1.55);
    \fill[blue!10] (0,0) circle (1.28);
    \draw[blue!70!black, line width=1.1pt]          (0,0) circle (1.28);

    \draw[refinearrow] (1.52,0.00) -- (1.32,0.00);

    \draw[-{Stealth[length=1mm,width=2mm]}, thick] (-1.95,-2.60) -- (1.95,-2.60) node[right] {shell index $m$};
    \foreach \x/\lbl in {-1.35/$m_L$, 0.00/$m^*$, 1.35/$m_U$}{
        \draw (\x,-2.50) -- (\x,-2.70);
        \node[tinylabel, below] at (\x,-2.72) {\lbl};
    }
    \node[circle, fill=blue!70!black, inner sep=1.7pt] at (0,-2.60) {};
    \draw[-{Stealth[length=1mm,width=2mm]}, blue!70!black, line width=0.8pt]
        (0.4,-1.21) to[out=210,in=28] (0.0, -2.45);
\end{scope}

\end{tikzpicture}
\caption{\small Schematic illustration of the global adaptive search for the infrared cutoff. Left: $S_{N_k}(\mbf q)$ is sampled on low-$q$ shells inside the BZ up to a prescribed cap $q_{\max}$. Middle: candidate infrared regions $ |\mbf q|\le q_t $ are expanded outward and tested for slope stability; the first unstable region defines an invalid cutoff. Right: the resulting valid--invalid bracket is refined to determine the largest stable cutoff $q_t^*=m^*q_{\min}$.\normalsize}
\label{fig:global_search_spheres}
\end{figure}
\paragraph{Global adaptive search.}
In the global variant, each candidate cutoff is assessed using all sampled points with $q\le q_t$. We first compute $S^{(\btheta)}_{N_k}(\mbf q)$ for all distinct momentum shells up to a prescribed cap $q_{\max}=q_t(m_{\rm cap})$. Starting from a minimum shell index $m_{\min}$, we then enlarge the fitting window by repeated doubling, refit $\alpha(m)$ at each stage, and stop once the stability criterion in Eq.~\eqref{eq:valid_predicate} fails. This produces a valid--invalid bracket $[m_L,m_U]$, which is then refined by binary search to determine the largest stable cutoff $m^*$. If the predicate does not fail before reaching $m_{\rm cap}$, double $m_{\rm cap}$, update the lower bound $m_L$ to $m_{\rm cap}$ and continue as above until a failure occurs. The output is the pair $(\alpha^*,q_t^*)$ computed by fitting against all unique data-points ($\eta_{k}^{\frac{2}{3}}$ of these) lying in the infrared region defined by $q_{t}^{*}$. In the worst case, the number of sampled shells required by the global search scales as $\mathcal O(N_k^{\frac{2}{3}})$. See Fig. \ref{fig:global_search_spheres} for a clear pictorial description of this algorithm.
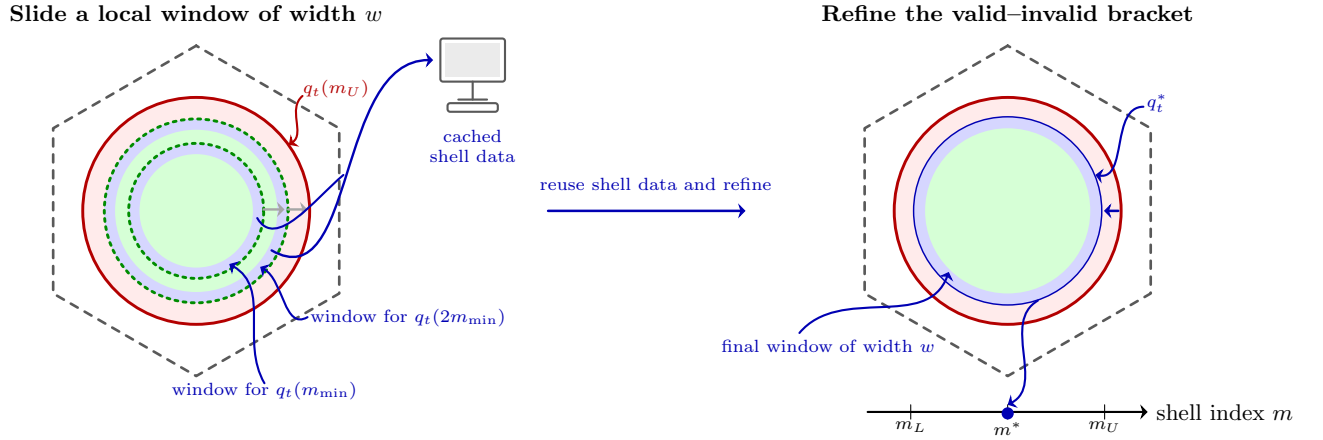
\begin{figure}[htbp]
\centering
\begin{tikzpicture}[
    scale=0.95,
    transform shape,
    line cap=round,
    line join=round,
    >=Latex,
    font=\small
]

\tikzset{
    bz/.style={draw=black!65, dashed, line width=1.0pt},
    invalidball/.style={draw=red!70!black, fill=red!8, line width=1.1pt},
    optball/.style={draw=blue!70!black, fill=blue!10, line width=1.1pt},
    corefill/.style={fill=green!16},
    windowfill/.style={fill=blue!16},
    qtcircle/.style={draw=green!55!black, dotted, line width=1.1pt},
    panelarrow/.style={-{Stealth[length=1mm,width=2mm]}, line width=0.9pt, gray!70},
    refinearrow/.style={-{Stealth[length=1mm,width=2mm]}, line width=0.95pt, blue!70!black},
    cachearrow/.style={-{Stealth[length=1mm,width=2mm]}, line width=0.9pt, blue!70!black},
    lab/.style={font=\small\bfseries},
    tinylabel/.style={font=\scriptsize}
}

\node[lab] at (3.6,5.0) {Slide a local window of width $w$};
\node[lab] at (14.9,5.0) {Refine the valid--invalid bracket};

\begin{scope}[shift={(3.6,2.25)}]
    \draw[bz] (90:2.3) \foreach \a in {150,210,270,330,30} { -- (\a:2.3)} -- cycle;

    \fill[red!8] (0,0) circle (1.58);
    \draw[red!70!black, line width=1.1pt] (0,0) circle (1.58);

    \fill[corefill] (0,0) circle (1.28);

    \fill[windowfill, even odd rule] (0,0) circle (1.28) (0,0) circle (1.13);

    \fill[windowfill, even odd rule] (0,0) circle (0.94) (0,0) circle (0.79);

    \draw[qtcircle] (0,0) circle (0.94);
    \draw[qtcircle] (0,0) circle (1.28);

    \draw[panelarrow] (0.94,0.02) -- (1.23,0.02);
    \draw[panelarrow] (1.28,0.02) -- (1.55,0.02);

    \node[tinylabel, blue!70!black] at (0.95,-2.50) {window for $q_t(m_{\min})$};
    \draw[-{Stealth[length=1mm,width=2mm]}, blue!70!black, line width=0.8pt]
        (0.95,-2.40) to[out=120,in=330] (0.46,-0.74);

    \node[tinylabel, blue!70!black] at (2.95,-1.48) {window for $q_t(2m_{\min})$};
    \draw[-{Stealth[length=1mm,width=2mm]}, blue!70!black, line width=0.8pt]
        (1.55,-1.48) to[out=210,in=330] (0.89,-0.78);
    \node[tinylabel, red!70!black] at (1.95,1.70) {$q_t(m_U)$};
    \draw[-{Stealth[length=1mm,width=2mm]}, red!70!black, line width=0.8pt]
        (1.48,1.60) to[out=220,in=35] (1.29,0.92);

    \begin{scope}[shift={(3.85,1.6)}]
        \draw[black!65, line width=0.8pt, rounded corners=1pt] (-0.45,0.20) rectangle (0.45,0.80);
        \fill[black!8] (-0.38,0.27) rectangle (0.38,0.73);
        \draw[black!65, line width=0.8pt] (0,0.20) -- (0,-0.05);
        \draw[black!65, line width=0.8pt] (-0.20,-0.05) -- (0.20,-0.05);
        \draw[black!65, line width=0.8pt, rounded corners=1pt] (-0.35,-0.22) rectangle (0.35,-0.10);
    \end{scope}

    \node[tinylabel, blue!70!black, align=center] at (3.85,0.9) {cached\\shell data};

    \draw[-, blue!70!black, line width=0.8pt]
        (0.85,-0.10) to[out=300,in=220] (2.05,0.50);
    \draw[cachearrow]
        (1.05,-0.55) to[out=330,in=180] (3.30,2.10);
\end{scope}

\draw[refinearrow] (8.50,2.25) -- (11.25,2.25);
\node[tinylabel, blue!70!black] at (10.00,2.62) {reuse shell data and refine};

\begin{scope}[shift={(14.9,2.25)}]
    \draw[bz] (90:2.3) \foreach \a in {150,210,270,330,30} { -- (\a:2.3)} -- cycle;

    \fill[red!8] (0,0) circle (1.58);
    \draw[red!70!black, line width=1.1pt] (0,0) circle (1.58);

    \fill[corefill] (0,0) circle (1.30);
    \draw[blue!70!black, line width=1.1pt] (0,0) circle (1.30);

    \fill[windowfill, even odd rule] (0,0) circle (1.30) (0,0) circle (1.15);

    \draw[refinearrow] (1.54,0.00) -- (1.34,0.00);

    \node[tinylabel, blue!70!black] at (2.08,1.50) {$q_t^*$};
    \draw[-{Stealth[length=1mm,width=2mm]}, blue!70!black, line width=0.8pt]
        (1.89,1.45) to[out=180,in=-8] (1.22,0.50);

    \node[tinylabel, blue!70!black] at (-2.5,-1.88) {final window of width $w$};
    \draw[-{Stealth[length=1mm,width=2mm]}, blue!70!black, line width=0.8pt]
        (-2.9,-1.70) to[out=50,in=250] (-0.82,-0.85);

    \draw[-{Stealth[length=1mm,width=2mm]}, thick] (-1.95,-2.80) -- (1.95,-2.80) node[right] {shell index $m$};
        \foreach \x/\lbl in {-1.35/$m_L$, 0.00/$m^*$, 1.35/$m_U$}{
            \draw (\x,-2.70) -- (\x,-2.90);
            \node[tinylabel, below] at (\x,-2.82) {\lbl};
        }
    \node[circle, fill=blue!70!black, inner sep=1.7pt] at (0,-2.82) {};

    \draw[-{Stealth[length=1mm,width=2mm]}, blue!70!black, line width=0.8pt]
        (0.42,-1.25) to[out=210,in=28] (0.00,-2.70);
\end{scope}

\end{tikzpicture}
\caption{\small Schematic illustration of the local windowed search for the infrared cutoff. Left: for each candidate cutoff $q_t(m)$, the infrared slope is estimated from a moving window of width $w$ over the most recent momentum shells (green), and only newly required shells are added as the window advances (blue). Previously computed shell data are retained and reused. Right: once a valid--invalid bracket is identified, the largest stable cutoff $q_t^*=m^*q_{\min}$ is obtained by refinement, while the final slope estimate is still taken from a local window of width $w$.\normalsize}
\label{fig:local_windowed_search}
\end{figure}
\paragraph{Local windowed search.}
In the local variant, the infrared slope is monitored through a moving window over nearby momentum shells rather than through a cumulative fit over all points below $q_t$. For a window of width $w$, we form a weighted local estimate of the slope using the most recent $w$ shells,
\begin{equation}
    \alpha_{\rm win}(m)
    =
    \frac{\sum_{j=m-w+1}^{m}\omega_j\,\alpha_j^{\rm(pt)}}{\sum_{j=m-w+1}^{m}\omega_j},
    \qquad
    \omega_j=\frac{1}{(\sigma_j^{\rm(pt)})^2},
\end{equation}
with the corresponding propagated uncertainty obtained from the same weights. Starting from an initial shell index $m_{\min}\ge w$, we again enlarge the candidate cutoff by doubling, but evaluate the local slope only on newly required shells. Once a valid--invalid bracket is found, the largest stable cutoff is determined by binary search exactly as in the global procedure. The output is again $(\alpha^*,q_t^*)$. Because previously computed shells are reused throughout the search, the total number of sampled shells scales as $\mathcal O(\frac{2}{3}w\log N_k)$. A pictorial description is provided by Fig. \ref{fig:local_windowed_search}.

\paragraph{Common classical post-processing.}
Once $(\alpha^*,q_t^*)$ have been determined by either search strategy, the finite-size correction is evaluated as
\begin{equation}
    \Delta V^{*}
    =
    \frac{\alpha^{*}}{4\pi^2}\int_{q\le q_t^{*}} dq\, (q^{\beta+2}-1)v(q)
    -
    \frac{1}{2\Omega\;M}\sum_{\btheta}
    \sum_{\mbf{q}_i;\,q_i\le q_t^{*}} v(q_i)\,\left(S^{(\btheta)}_{N_k}(\mbf{q}_i)-1\right)~.
\end{equation}

\paragraph{Practical trade-off.}
The global search is more direct and typically more robust because each candidate cutoff is assessed using the full set of points below it, but this can be more cost intensive. The local search is more sample-efficient because it reuses previously computed shell values, but it requires additional classical storage and bookkeeping. In both cases, the procedure returns the same physically relevant quantities: a stable infrared window and the corresponding slope parameter that determines the leading finite-size correction.
\subsection{A quantum algorithm for estimating the static structure factor}
\label{sec: quantum_algorithm}

The finite-size correction developed in our work requires accurate values of the SSF at a small set of low-momentum points. On a fault-tolerant quantum computer, these quantities are naturally accessed as expectation values in the ground state as given in Eq. \eqref{eq: SSF_def_twisted}. The purpose of this subsection is to show how the problem can be reduced to estimating a single-qubit success probability, and to state the corresponding resource requirements. This step is central to the paper: once the low-$q$ values of $S^{(\btheta)}_{N_k}(\mbf q)$ are available fault tolerantly, they can be combined with the infrared reconstruction procedure of the previous subsection to produce the leading two-body finite-size correction. 
\\

We assume access to two ingredients. First, let $U_{\hat S}$ be a $(\lambda_{\hat S},m,0)$ block-encoding of $\hat S_{\mbf q}$, so that
$
    \hat S_{\mbf q}
    =
    \lambda_{\hat S}\,
    (\langle 0^m|\otimes I)\,
    U_{\hat S}\,
    (|0^m\rangle\otimes I)~.
$
Second, let $U_\psi$ prepare a trial state $|\psi^{(\btheta)}\rangle$ with nonzero overlap with the true ground state,
$
    U_{\psi}\,|0^k\rangle = |\psi^{(\btheta)}\rangle,
$ with $
    |\langle \psi_0^{(\btheta)}|\psi^{(\btheta)}\rangle|^2 = p_0~.
$
The block-encoding provides coherent access to the observable $\hat S_{\mbf q}$, while the state-preparation oracle supplies the many-body state on which the observable is to be evaluated. The key observation is that a Hadamard test converts the desired expectation value into the probability of obtaining a single measurement outcome on a control qubit. In circuit form, 
\begin{center} 
\begin{quantikz} \lstick{$\ket{0}$} & \gate[wires=3]{U_{\varphi}} & \meter{} \\ \lstick{$\ket{0^{m}}$} & & \qw\\ \lstick{$\ket{0^{k}}$} & & \qw \end{quantikz} $\;=\;$ \begin{quantikz} \lstick{$\ket{0}$} & \gate{H} & \ctrl{1} & \gate{H} & \meter{} \\ \lstick{$\ket{0^{m}}$} & \qw & \gate[wires = 2]{U_{\hat{S}}} & \qw & \qw\\ \lstick{$\ket{0^{k}}$} & \gate{U_{\psi}} & & \qw & \qw 
\end{quantikz} 
\end{center} 
where the ancilla register of size $m$ is required by the block-encoding, while the $k$-qubit system register stores the prepared trial state.
Denoting the corresponding state-preparation circuit by $U_\varphi$, one obtains
\begin{align}
    U_{\varphi}\ket{0}\ket{0^{m}}\ket{0^{k}}
    &=
    |\varphi\rangle \nonumber\\
    &=
    \frac{1}{2}|0\rangle\,(I+U_{\hat S})\,|0^m\rangle|\psi^{(\btheta)}\rangle
    +
    \frac{1}{2}|1\rangle\,(I-U_{\hat S})\,|0^m\rangle|\psi^{(\btheta)}\rangle~\nn.
\end{align}
The success event is the measurement outcome $|0\rangle$ on the control qubit. Its probability
in the ideal case $|\psi^{(\btheta)}\rangle=|\psi^{(\btheta)}_0\rangle$, reduces to
\begin{equation}
    P(0)
    =
    \frac{1}{2}
    \left(
    1+\frac{S^{(\btheta)}_{N_k}(\mbf q)}{\lambda_{\hat S}}
    \right).
\end{equation}
Estimating $S^{(\btheta)}_{N_k}(\mbf q)$ is therefore equivalent to estimating the single probability $P(0)$. At this point one could estimate $P(0)$ by repeated sampling of the Hadamard-test circuit. For the resource estimates relevant here, however, it is more useful to combine the Hadamard test with amplitude amplification and phase estimation, which reduce the cost of amplitude estimation from the usual Monte Carlo sampling scaling to Heisenberg-like scaling of $\mathcal{O}(\epsilon_P^{-1})$. The corresponding construction is standard and is summarized in Appendix~\ref{app: amplitude_amplification}; here we retain only the resulting scaling needed later in the paper. Specifically, an estimate of $S^{(\btheta)}_{N_k}(\mbf q)$ to additive precision $\epsilon_S$ can be obtained with overall query complexity
$\mathcal{O}\!\left(p_0^{-1/2}\,\epsilon_P^{-1}\right),$ where $\epsilon_P$ is the precision with which the Hadamard-test success probability is resolved. If the state-preparation error is denoted by $\epsilon_\psi$, a convenient choice as per Lemma 7 in Appendix D of \cite{Tong_2021} is $\epsilon_P = \frac{\epsilon_S}{2\lambda_{\hat S}\log(\epsilon_\psi^{-1})}~.$
\\

The corresponding logical resource requirements per oracle call are
\begin{align}
    \text{Qubit count}
    &=
    k+m+1+\left\lceil \log\!\left(p_0^{-1/2}\epsilon_P^{-1}\right)\right\rceil~,\\
    \text{Toffoli count per query}
    &=
    T_{U_\psi}
    +T_{U_{\hat S}}
    +T_{U_\psi^\dg}
    +T_{U_{\hat S}^\dg}
    +T_{\rm REF}~\label{eq: quantum_algorithm_resource}.
\end{align}
Here the additional $1$ is the Hadamard-test control qubit, and the final register size accounts for the phase-estimation ancillas needed to resolve $P(0)$ to the required precision. The Hadamard and $-Z$ gates are Clifford operations and therefore do not contribute to the leading Toffoli count.
\\

The cost of state preparation merits separate comment. In insulating systems, available strategies suggest that the cost of preparing a sufficiently good trial state may remain subleading relative to the dominant block-encoding cost \cite{Lin_notes, Marteau_2023, Dong_2022, Fomichev_2024, Pathak_2023}. In particular, block-encoding-based preparation can inherit the qubit scaling of the underlying Hamiltonian encoding, and compact multi-determinant representations may reduce the Toffoli count further when the number of important determinants is small \cite{Fomichev_2024}. Metallic systems are more challenging because the small spectral gap and potentially large wavefunction complexity can substantially increase the cost of preparing a useful starting state. At present, no uniformly optimal state-preparation strategy is known for the full range of material classes considered here. Nevertheless, this overhead is not unique to the present protocol: any fault-tolerant ground-state energy calculation must also prepare an approximate ground state. Since the main purpose of our work is to show that estimating the low-$q$ SSF is subleading relative to the underlying energy-estimation routine, the state-preparation cost can be treated as a shared overhead and discussed separately from the incremental cost of the present method.
\\

Taken together, these observations establish the main algorithmic point of this section. The Hadamard test converts $S^{(\btheta)}_{N_k}(\mbf q)$ into a single-qubit success probability, and amplitude amplification then allows that probability to be estimated with the precision needed for the finite-size correction. The remainder of the paper uses these resource expressions to compare the cost of evaluating $\Delta V$ with that of the underlying ground-state energy calculation.
\section{Results}
\label{sec: results}
\subsection{Noise tolerance of the static structure factor computation}
\label{sec:noise_tolerance_main}

We now quantify the precision with which the low-$q$ values of $S(\mbf q)$ must be estimated in order for the finite-size correction $\Delta V$ to be useful at the target accuracy $\epsilon$. The main point of this section is that the correction routine is considerably more noise tolerant than the underlying ground-state energy estimation itself. In other words, the values of $S(\mbf q)$ entering the infrared reconstruction do not need to be computed to the same absolute precision as the total energy. This separation of scales is what allows the SSF route to finite-size mitigation to remain fault-tolerantly affordable.
\\

Our starting point is to parameterize the uncertainty in the computed structure factor as $\sigma_S = p\,a\,\epsilon$, where $a=(\Omega_{\rm unit})^{1/3}$ and $p$ is a dimensionless prefactor that measures how much larger the tolerable noise in $S(\mbf q)$ may be relative to the final target accuracy $\epsilon$. As shown in Appendix~\ref{app:noise_tolerance_details}, we show that $p$ can grow with system size without spoiling the accuracy of $\Delta V$: once the error budget is propagated through both the fitted continuum contribution and the discrete low-$q$ sum, one finds the admissible bound
\begin{equation}
    p \leq \mathfrak{p}N_k^{\frac{1}{3}}~.\label{eq:p_bound}
\end{equation}
Here the constant prefactor $\mathfrak{p}$, defined more explicitly in Appendix \ref{app:noise_tolerance_details}, is determined by the more restrictive of the two error channels: the continuum infrared integral and the discrete shell sum. The important consequence is that the allowable noise in $S(\mbf q)$ grows with the linear extent of the BvK cell. Thus, although $\Delta V$ is essential for obtaining the thermodynamic-limit energy, its ingredients do not need to be estimated at the same stringent precision as the total energy itself.
\\

This looser precision requirement has a direct algorithmic consequence. The fault-tolerant cost of the correction routine is set by the number of low-$q$ structure-factor evaluations required by the cutoff-search procedure, multiplied by the cost of estimating each such value through the Hadamard-test-based protocol described previously. Combining the noise bound in Eq.~\eqref{eq:p_bound} with the global and local search strategies of Section~\ref{sec: infrared_behavior_cutoff_procedure}, one obtains
\begin{equation}
    \text{Toffoli-gate count for} ~\Delta V\sim \begin{cases}
        \mathcal{O}\!\left(\sqrt{M}\frac{N_k^{\frac{2}{3}}}{p\epsilon}\,
    \lambda_{U_{\hat{S}}}\,
    T_{U_{\hat{S}}}\right)~, & \text{Global binary search}\\
    \mathcal{O}\!\left(\sqrt{M}\frac{\max{\{\eta_k^{\frac{2}{3}},\frac{2}{3}\log {N_k}}\}\,}{p\epsilon}
    \lambda_{U_{\hat{S}}}\,
    T_{U_{\hat{S}}}\right)~, & \text{Local binary search}~.\label{eq: toffoli_deltaV_general}
    \end{cases}
\end{equation}
 We have also absorbed the lattice constant $a$ into the one-norm so that it has the same dimensions as that of the energy.  The role of $p$ is transparent: because the admissible noise level in $S(\mbf q)$ is parametrically larger than $\epsilon$, the required precision per structure-factor evaluation is correspondingly weaker than that of the main energy routine. This is precisely why the cost of evaluating $\Delta V$ does not compete with the cost of computing the ground-state energy itself. As already discussed in Section \ref{sec: quantum_algorithm} we are ignoring the state preparation cost in the overall Toffoli count of the fault tolerant computation of $\Delta V$.
\\

The same analysis also clarifies the regime in which the infrared ansatz $S(\mbf q)\sim \alpha q^\beta$ is self-consistent at the target accuracy. Using Eq. \eqref{eq: N_k_constraint} and Eq.~\eqref{eq:p_bound} we get a renewed bound for $N_k$:
\begin{equation}
    N_k\gtrsim \left[\frac{(2\pi)^{3}}{\Omega_{\rm sc}}\right]^{\frac{\beta+1}{\beta+2}}
\left(\frac{\gamma}{\mathfrak{p}a\epsilon}\right)^{\frac{3}{\beta+2}}~.
\end{equation}
Once this regime is reached, the tolerance encoded in Eq.~\eqref{eq:p_bound} ensures that the structure-factor data may still be computed with a relaxed precision budget. The picture that emerges is therefore favorable. The correction $\Delta V$ is sensitive only to a small number of infrared modes, and the required accuracy of those modes is itself substantially less demanding than the accuracy required for the total energy. The SSF correction is thus not merely formally subleading; it is also operationally subleading in the fault-tolerant setting.
\subsection{Static structure factor in the Bloch basis}
\label{sec:bloch_basis_structure_factor}

We now recast the SSF in a Bloch-orbital representation suited to periodic materials calculations and, in later sections, to fault-tolerant block encoding. The goal of this section is twofold. First, we identify the momentum structure of the density operator in a basis natural for crystalline systems. Second, we show that the SSF retains a simple bilinear form, with all material-specific information compressed into unit-cell form factors. This separation between momentum selection rules and intra-cell physics is the key structural feature that makes the operator both physically transparent and algorithmically tractable.
\\

We begin from the Bloch expansion of the field operator,
\begin{equation}
    \hat{\psi}(\mbf{x}) = \frac{1}{\sqrt{N_k}}\sum_{\mbf{k}}^{N_k}\sum_{ i,\sigma}\psi_{(\mbf{k},i)}(\mbf{x})\,\hat{a}_{(\mbf{k},i),\sigma}~.
\end{equation}
Here $\mbf{k}$ runs over the $N_k$ points of the BZ mesh, $i$ labels the orbital or band index, and $\sigma$ denotes spin. The normalization reflects BvK boundary conditions on a finite $k$-mesh and a simulation cell containing $N_a$ atoms. To make translational symmetry explicit, we write each Bloch orbital in the standard form
\begin{equation}
    \psi_{(\mbf{k},i)}(\mbf{r}) = e^{i\mbf{k}\cdot \mbf{r}}\,u_{(\mbf{k},i)}(\mbf{r}),
\end{equation}
where $u_{(\mbf{k},i)}(\mbf{r})$ is periodic on the unit cell. Once this decomposition is inserted into the Fourier-space density operator, the lattice sum enforces crystal-momentum conservation up to a reciprocal lattice vector. As a result, the density operator couples only orbital pairs whose crystal momenta differ by a transfer momentum $\mbf Q$, modulo an Umklapp vector $\mbf G$. This structure allows the density operator at momentum $\mbf q=\mbf Q+\mbf G$ to be written in the compact form
\begin{equation}
    \hat{\rho}_{\mbf{Q}+\mbf{G}} = \sum_{\mbf{k}}^{N_k}\sum_{i,j}^{N_b/2}
    F_{\mbf{k}i,\mbf{k}\oplus\mbf{Q}j}(\mbf{G})~\hat{E}_{\mbf{k}i,\mbf{k}\oplus\mbf{Q}j}~.
\end{equation}
The fermionic excitation operators are
$
    \hat{E}_{\mbf{k}i,\mbf{k}'j} \;\equiv\; \sum_{\sigma}\hat{a}^{\dg}_{(\mbf{k},i),\sigma}\hat{a}_{(\mbf{k}',j),\sigma},
$
and the corresponding form factors are
\begin{equation}
    F_{\mbf{k}i,\mbf{k}\oplus\mbf{Q}j}(\mbf{G}) = 
    \int_{\Omega_{\rm sc}}d^3\mbf{r}~
    e^{i\mbf{G}\cdot\mbf{r}}
    u^{*}_{(\mbf{k}\oplus\mbf{Q},i)}(\mbf{r})\,u_{(\mbf{k},j)}(\mbf{r})~.
\end{equation}
 These form factors contain the entire supercell-scale information of the problem: orbital character, interband couplings, and local-field effects all enter through $F$. By contrast, the dependence on $\mbf Q$ and $\mbf G$ expresses the momentum-selection structure imposed by translation symmetry. In a plane-wave basis this distinction collapses to a simple selection rule, whereas in a general Bloch basis it remains nontrivial and physically informative. With this notation, the SSF operator takes the familiar density-density form
\begin{equation}
    \hat{S}_{\mbf{Q}+\mbf{G}} \;=\;  \hat{\rho}^{\dg}_{\mbf{Q}+\mbf{G}}\,\hat{\rho}_{\mbf{Q}+\mbf{G}}~,
\end{equation}
which shows that the operator of interest is completely determined once $\hat\rho_{\mbf Q+\mbf G}$ is known. In the Bloch basis, the SSF is still a quadratic operator in the density, but the density itself is now resolved into momentum-conserving inter-orbital excitations weighted by material-dependent form factors. The decomposition is therefore both physically natural and directly compatible with operator constructions based on linear combinations of fermionic excitations.
\\

Several consequences are worth emphasizing. First, long-wavelength finite-size corrections probe only a small subset of transfer momenta. In the infrared regime relevant for $\Delta V$, one typically requires only the smallest values of $|\mbf q|$, which in practice strongly restricts the allowed pairs $(\mbf Q,\mbf G)$. This means that $\mbf G=0$ and only a small region of the BZ contributes. Second, all information from the underlying electronic structure enters through the precomputed form factors $F$, while the operator algebra is carried by the excitation operators $\hat E$. Third, this representation makes clear that the low-$q$ sector relevant to finite-size corrections does not require access to the full Hamiltonian structure. Instead, it is enough to target a restricted set of transfer-momentum sectors.
\subsection{The density operator as a linear combination of unitaries}
\label{sec:lcu_density_main}

To estimate $S(\mbf q)$ using the framework of Section~\ref{sec: quantum_algorithm}, we require a block encoding of the SSF operator $\hat S_{\mbf q}$. The central observation is that $\hat S_{\mbf q}$ can be built directly from a block encoding of the density operator $\hat\rho_{\mbf q}$. In particular, once a block encoding $U_{\hat\rho}$ is available, a block encoding of $\hat S_{\mbf q}=\hat\rho_{\mbf q}^{\dg}\hat\rho_{\mbf q}$ is obtained through the reflected product
$
    U_{\hat{S}}
    =
    [U_{\hat{\rho}}]^{\dg}
    \cdot
    \mathrm{REF}
    \cdot
    U_{\hat{\rho}}
$
with the corresponding circuit that looks as follows: 
\begin{center} 
\begin{quantikz} \lstick{$\ket{0}$} & \gate[wires=2]{U_{\hat{S}}} & \qw \\ \lstick{$\ket{0^{k\;+\;m}}$} & & \qw \end{quantikz} $\;=\;$ \begin{quantikz} \lstick{$\ket{0}$} & \gate[wires=2]{U_{\hat{\rho}}} & \gate[wires=2]{\mathrm{REF}} & \gate[wires=2]{U^{\dg}_{\hat{\rho}}} & \qw \\ \lstick{$\ket{0^{k\;+\;m}}$} & & & & \qw \end{quantikz} 
\end{center}
Here $\mathrm{REF}$ denotes the standard reflection about the all-zero ancilla state. This construction shows immediately that the leading cost of block encoding $\hat S_{\mbf q}$ is inherited from the cost of block encoding $\hat\rho_{\mbf q}$, while the additional reflection and Hermitian conjugation contribute only subleading overhead. The corresponding normalization factor is determined by the one-norm of the LCU representation of $\hat\rho_{\mbf q}$. Specifically,
\begin{equation}
    \lambda_{\hat{S}} = \frac{1}{2}\left(\lambda_{\rho}\right)^2~.\label{eq: onE_{n}orm_structure_factor}
\end{equation}
Thus, the problem reduces to finding a useful LCU decomposition of the density operator together with its associated one-norm. The construction we use proceeds by decomposing the form-factor tensor in each $\mbf k$-sector into Hermitian components, diagonalizing the resulting single-particle matrices, and then expressing the rotated number operators in terms of Pauli $Z$ operators. The result is the following LCU form:
\begin{equation}
    \hat{\rho}_{\mbf{q}}=
    \frac{1}{2}\sum_{J=0,1}\sum_{\mbf{k}}^{N_k}\sum_{i=1}^{N_b/2}\sign(f^{(J)}_{i})|
    f^{(J)}_{i}(\mbf{q},\mbf{k})|\;
    \hat{\mbf{U}}^{(J)}_{\mbf{k}}(\mbf{q})\;
    Z^{(J)}_{\mbf{k},i}\;
    \hat{\mbf{U}}^{(J)\dg}_{\mbf{k}}(\mbf{q})~,
\end{equation}
with one-norm
\begin{equation}
    \lambda_{\rho}
    =
    \frac{1}{2}\sum_{J=0,1}\sum_{\mbf{k}}^{N_k}\sum_{i=1}^{N_b/2}
    \left|f^{(J)}_{i}(\mbf{q},\mbf{k})\right|~.
\end{equation}
The index $J\in\{0,1\}$ labels the two Hermitian components of the non-Hermitian form-factor tensor, $\hat{\mbf U}^{(J)}_{\mbf k}(\mbf q)$ denotes the Givens-rotation unitary that diagonalizes the corresponding single-particle matrix in the $\mbf k$-sector, and $f_i^{(J)}(\mbf q,\mbf k)$ are the resulting eigenvalues. The operators $Z^{(J)}_{\mbf k,i}$ are Pauli $Z$ operators acting on the rotated fermionic modes. In other words, all material-specific information is compressed into the coefficients $f_i^{(J)}$ and the corresponding orbital rotations $\hat{\mbf U}^{(J)}_{\mbf k}(\mbf q)$, while the operator algebra itself reduces to a weighted sum of conjugated Pauli strings. A complete derivation of these expressions is given in Appendix~\ref{app:lcu_density_details}.
\\

This form makes the asymptotic scaling transparent. The eigenvalues $f_i^{(J)}$ remain $\mathcal O(1)$ as the basis is enlarged, then $\lambda_\rho$ grows linearly with the number of Bloch-orbital modes, i.e. as $\mathcal O(N_bN_k)$. Equation~\eqref{eq: onE_{n}orm_structure_factor} then implies that $\lambda_{\hat S}$ scales quadratically, 

\begin{equation}
   \lambda_{\hat S}\sim \mathcal O((N_bN_k)^2)~. \label{eq: asymptotic_lambda}
\end{equation}
As a result, the query complexity of the SSF estimation inherits the same polynomial dependence on $N_bN_k$ as the main energy-estimation routine, but without an explicit dependence on $N_{\rm pw}$, since the present construction is targeted directly at the infrared density response. This is the key structural reason why the evaluation of $\Delta V$ can remain subleading even though it probes a genuinely many-body observable.
\subsection{Block encoding of the density operator}
\label{sec:block_encoding_density_main}
We now describe how the density operator $\hat\rho_{\mbf q}$ is block encoded in the Bloch-orbital basis. This construction is the core of the fault-tolerant algorithm: once a block encoding of $\hat\rho_{\mbf q}$ is available, the block encoding of the SSF follows directly from the construction of Section~\ref{sec:lcu_density_main}. In the main text we focus on the general case away from the $\Gamma$-point, since it contains the full lattice-momentum sector structure relevant to periodic materials. The $\Gamma$-point construction and the complete resource formulas are deferred to Appendix~\ref{app:block_encoding_density_details}. 
\\

The block encoding is implemented as a walk operator of the standard prepare--select--unprepare form~\cite{Lin_notes,Babbush_2018},
\begin{equation}
    \hat{\mathcal{Q}}^{(\mbf{q})}_{W} = \PREP_{\mbf{q}}\cdot\SEL_{\mbf{q}}\cdot\PREP^{\dg}_{\mbf{q}}~.
\end{equation}
At a high level, $\PREP_{\mbf q}$ prepares the LCU superposition over the terms appearing in the density-operator decomposition, including their signs and momentum labels, while $\SEL_{\mbf q}$ applies the corresponding controlled one-body transformation on the appropriate $(\mbf k,\mbf k\oplus \mbf q)$ sector. Relative to the $\Gamma$-point case, the essential additional ingredient is the explicit treatment of lattice-momentum sector labels: the circuit must not only load the appropriate Givens-rotation data, but also route the action of the selected unitary to the correct pair of momentum sectors. This is accomplished by a classical-quantum adder for the map $\mbf k\mapsto \mbf k\oplus \mbf q$ together with momentum- and spin-controlled swap operations, following the general block-encoding framework of Refs.~\cite{Babbush_2018,Rubin_bloch_2023,Gidney_2025}.
\begin{figure}[ht!]
    \centering
    \small
    \begin{quantikz} 
    \lstick{succ $\ell$} & \gate[wires=2]{\text{prep}_{\ell}} & \qw & \ctrl{2} & \qw & \qw & \ctrl{6} & \qw & \qw  &\qw& \qw & \gate[wires=2]{\text{prep}^{\dg}_{\ell}} & \qw \\ \lstick{$\ell$} & \qw & \gate{\text{In}_{\ell}} & \qw & \qw & \qw & \qw & \qw & \qw & \qw & \gate{\text{In}_{\ell}} & \qw & \qw
    \\ \lstick{\sign} & \qw & \gate[wires=3]{{\rm data}_{\ell}} \vqw{-1} & \gate{Z} & \qw & \qw & \qw & \qw & \qw &\qw & \gate[wires=3]{{\rm data}_{\ell}} \vqw{-1} & \qw & \qw \\
    \lstick{$\mbf{Q}$} & \qw &  & \gate{\hat{A}_{\oplus \mbf{q}}} & \ctrl{3} & \qw & \qw & \qw & \ctrl{3} & \gate{\hat{A}_{\ominus \mbf{q}}} &  & \qw & \qw \\
    \lstick{rotations} & \qw & & \qw & \qw & \ctrl{2} & \qw & \ctrl{2} & \qw & \qw & & \qw & \qw \\ \lstick{spin} & \gate{H} & \qw & \ctrl{1} & \qw & \qw & \qw & \qw & \qw & \ctrl{1} & \qw & \gate{H} & \qw \\ \lstick{$|\psi^{(\btheta)}_{\downarrow}\rangle$} & \qw & \qw & \swap{1} & \targX{} & \gate{R} & \gate{Z_1} & \gate{R^{\dg}} & \targX{} & \swap{1} & \qw & \qw & \qw \\ \lstick{$|\psi^{(\btheta)}_{\uparrow}\rangle$} & \qw & \qw & \targX{} & \qw & \qw & \qw & \qw & \qw & \targX{} & \qw & \qw & \qw 
    \end{quantikz}
    \normalsize
    \caption{\small Quantum circuit for the block encoding of the density operator away from the $\Gamma$-point. The circuit implements the walk operator $\hat{\mathcal Q}^{(\mbf q)}_W=\PREP_{\mbf q}\cdot\SEL_{\mbf q}\cdot\PREP_{\mbf q}^{\dg}$. Relative to the $\Gamma$-point construction, the principal new ingredient is the explicit treatment of lattice-momentum sector labels. The $\PREP_{\mbf q}$ stage prepares the LCU index, sign, and momentum registers, while the $\SEL_{\mbf q}$ stage loads the corresponding Givens-rotation data through $\mathrm{In}_{\ell}$ and $\mathrm{data}_{\ell}$, applies the classical-quantum adder $\hat A_{\oplus\mbf q}$ to generate the shifted momentum index $\mbf k\oplus\mbf q$, and performs the required wave-vector- and spin-controlled swap operations so that the sequence $R\,Z_1\,R^\dg$ acts on the correct $(\mbf k,\mbf k\oplus\mbf q)$ sector. The inverse data loading and the final application of $\PREP_{\mbf q}^{\dg}$ uncompute the ancilla registers and complete the block encoding.\normalsize}
    \label{fig:rho_k_point_block_encoding}
\end{figure}
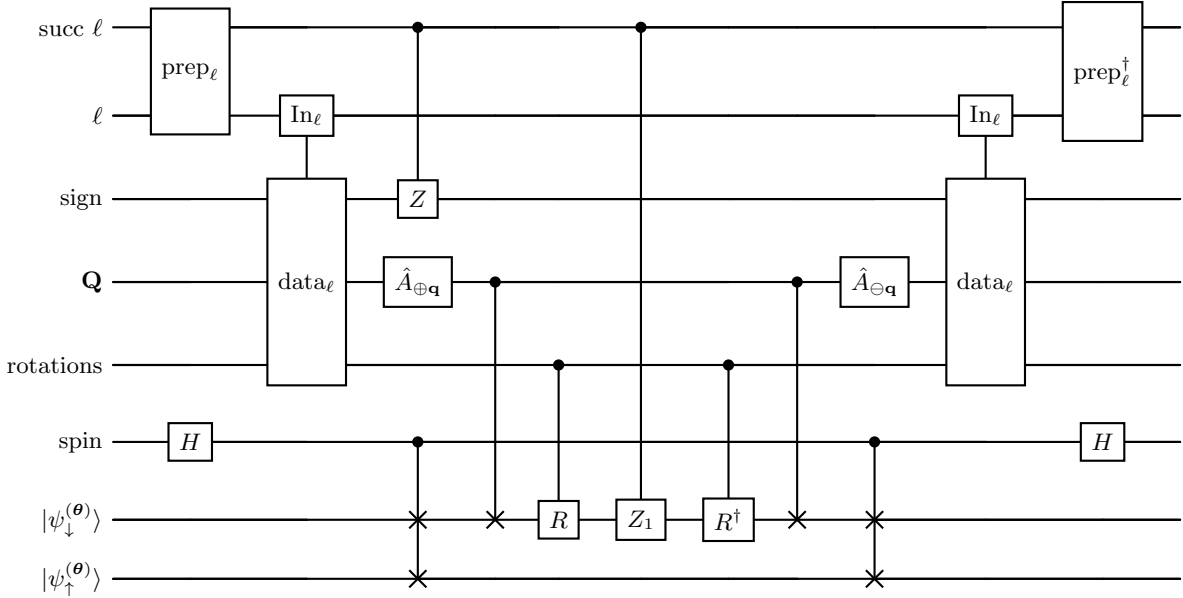
Fig.~\ref{fig:rho_k_point_block_encoding} gives the resulting circuit. The prepare stage encodes the LCU index and the amplitude weights associated with the selected term. The select stage then loads the relevant orbital-rotation data, constructs the shifted momentum label $\mbf k\oplus\mbf q$, and applies the controlled sequence $R\,Z_1\,R^\dg$ only on the selected momentum-spin sector. The subsequent uncomputation restores all ancilla registers to the reference state. The detailed formulas show that the dominant arithmetic and QROAM costs arise from loading rotation angles, manipulating momentum labels, and implementing controlled swaps across the $N_k$ sectors~\cite{Babbush_2018,Rubin_bloch_2023,Gidney_2025}. The key qualitative conclusion is that the general away-from-$\Gamma$ construction remains linear in the total number of Bloch-orbital momentum modes. More precisely, the explicit formulas derived in Appendix~\ref{app:block_encoding_density_details} imply
$
    T_{\hat{\rho}},~Q_{\hat{\rho}} \sim \mathcal{O}(N_bN_k)~.
$
Thus, although the general periodic case requires additional bookkeeping relative to the $\Gamma$-point construction, the asymptotic scaling of the density-operator block encoding remains linear in system size. Combining this with the quadratic scaling of the normalization factor derived in Section~\ref{sec:lcu_density_main} gives the contribution of the SSF block encoding to the overall algorithm,
\begin{align}
    T_{\hat S} &\sim \mathcal{O}\big((N_bN_k)^3\big)~,\label{eq: S_toffoli_count}\\
    Q_{\hat S} &\sim \mathcal{O}(N_bN_k)~\label{eq: S_qubit_count}.
\end{align}
The Toffoli count is cubic because the linear cost of the density-operator block encoding is multiplied by the quadratic normalization inherited from $\lambda_{\hat S}$, whereas the qubit count remains linear. This scaling is consistent with the qualitative picture developed throughout the paper: while the evaluation of the low-$q$ SSF is nontrivial, its cost remains parametrically controlled and subleading relative to that of the ground-state energy estimation routine. We discuss this comparison in more detail in Section~\ref{sec: cost_comparison}.
\\

The away-from-$\Gamma$ block encoding provides the general periodic implementation needed for the structure-factor algorithm. The construction is based on a standard qubitization quantum walk operator, augmented by lattice-momentum sector routing and arithmetic, and it retains linear asymptotic scaling in $N_bN_k$. Appendix~\ref{app:block_encoding_density_details} gives the full derivation, including the $\Gamma$-point specialization, explicit resource formulas, and the corresponding circuit-level decompositions.
\subsection{Cost comparison with energy estimation}
\label{sec: cost_comparison}

We now compare the cost of evaluating the finite-size correction $\Delta V$ with that of the underlying ground-state energy estimation routine. This comparison is the final step needed to place the present method in context. The central question is not whether the SSF calculation is itself inexpensive ---which it is evidently not, from our discussion so far --- but whether it remains subleading once incorporated into a full fault-tolerant workflow for periodic materials.
\\

The dominant cost of the $S(\mbf q)$ estimation routine arises from repeated applications of the reflection $R_\varphi$ used in amplitude amplification (for details see Appendix \ref{app: amplitude_amplification}). Combining the Hadamard-test construction given in Section \ref{sec: quantum_algorithm} and Eq. \eqref{eq: quantum_algorithm_resource} with the block encoding of $\hat S_{\mbf q}$ derived above gives the per-query cost
\begin{equation}
    T_{R_{\varphi}} = 4T_{U_{\hat{\rho}}}\;+\;3T_{\rm REF}\;+\;\mathcal{O}(\text{cost of state preparation})~.
\end{equation}
Here $T_{U_{\hat\rho}}$ is the cost of block encoding the density operator, while $T_{\rm REF}$ is the cost of the all-zero reflection. The latter scales only linearly in the size of the reflected register and is therefore subleading, since the ancilla registers entering the reflection grow only logarithmically with system size. At leading order, the cost per query is therefore set by the density-operator block encoding, with state preparation entering as an additional overhead shared with the energy-estimation problem itself. The overall cost of the finite-size correction follows by combining this per-query cost with the scaling of the normalization factor, $\lambda_{\hat S}\sim \mathcal{O}\!\big((N_bN_k)^2\big)$, and with the number of low-$q$ values of $S(\mbf q)$ required by the infrared reconstruction summarized in Section \ref{sec: infrared_behavior_cutoff_procedure} previously. Using Eq. \ref{eq: S_toffoli_count} and \ref{eq: asymptotic_lambda} along with \eqref{eq: toffoli_deltaV_general} we get the resulting total Toffoli complexity as
\begin{equation}
    T_{\Delta V}\sim \begin{cases}
        \mathcal{O}\!\left(\frac{\sqrt{M}}{\epsilon}\,N_k^{\frac{1}{3}}
    (N_bN_k)^3\right)~, & \text{Global binary search}~,\\[6pt]
        \mathcal{O}\!\left(\frac{\sqrt{M}}{\epsilon}\frac{\max{\{\eta_k^{\frac{2}{3}},\frac{2}{3}\log {N_k}\}}}{N_k^{\frac{1}{3}}}\,
    (N_bN_k)^3\right)~, & \text{Local binary search}~.
    \end{cases}
\end{equation}
The distinction between the two cases is physically transparent. In the global procedure, the cost reflects the need to explore a broader set of infrared shells before identifying the asymptotic regime. In the local procedure, previously computed shell data are reused, so the search overhead is reduced to the number of retained infrared modes $\eta_k$ together with a logarithmic dependence on $N_k$. The local strategy is therefore the more favorable one whenever the sliding-window analysis is sufficient to identify a stable infrared cutoff. To interpret these scalings, it is useful to compare them with the leading cost of fault-tolerant ground-state energy estimation for plane-wave Hamiltonians, which scales as $T_E\sim \mathcal{O}\!\big(\frac{\sqrt{M}}{N_k\epsilon}N_{\rm pw}^{\frac{3}{2}}(N_bN_k)^3\big)$ \cite{Rubin_bloch_2023, Bhardwaj:2026koe}. The important point is that both problems share the same dominant cubic dependence on $N_bN_k$; the difference lies in the prefactor multiplying that common scaling. For the finite-size correction, that prefactor is controlled by the number of infrared modes that must be sampled and by the search overhead used to identify the cutoff $q_t$. For the energy-estimation problem, by contrast, it is controlled by the full reciprocal space, resulting in a much larger plane-wave factor $N_{\rm pw}^{\frac{3}{2}}$ for the gate count. As a result, the cost of evaluating $\Delta V$ remains subleading even in the conservative global-search setting, provided
$N_{\rm pw}^{\frac{3}{2}}\gg N_k^{\frac{4}{3}}$. This separation is expected to hold for realistic periodic solids, where the plane-wave basis prefactor is typically much larger than the size of the BZ-mesh. In the local-search setting, the separation is even more pronounced because the search overhead is reduced to $\max\{\eta_k^{\frac{2}{3}},\frac{2}{3}\log N_k\}$. The same qualitative conclusion holds for logical qubits. The finite-size correction requires $Q_{\Delta V}=\mathcal{O}(N_bN_k)$, whereas the corresponding ground-state energy estimation routine scales as $Q_E=\mathcal{O}(N_{\rm pw}^{\frac{1}{2}}N_bN_k)$. Thus, the correction does not alter the leading qubit scaling of the overall simulation and adds only a subleading logical-memory overhead to the main calculation.
\\

This comparison sharpens the main message of the paper. The leading two-body finite-size correction in periodic quantum simulations can be extracted from a small set of low-momentum values of the SSF at a fault-tolerant cost that remains below that of the underlying ground-state energy estimation routine. In other words, finite-size mitigation need not be achieved by pushing the main many-body calculation to prohibitively dense $k$-meshes. By combining twist averaging to reduce one-body shell effects with an infrared reconstruction of $\Delta V$ from $S(\mbf q)$, one obtains a thermodynamic-limit correction strategy that is both physically well motivated and asymptotically compatible with fault-tolerant resource constraints. This is precisely what makes the SSF a useful target observable in the fault-tolerant setting: it captures the long-wavelength two-body physics that dominates the residual finite-size error, yet it can be accessed at a cost that does not overtake the main simulation itself.
\section{Conclusion}
Our work develops a fault-tolerant quantum post-processing strategy for finite-size mitigation in periodic many-body simulations. The key point is that, after one-body shell effects are suppressed by twist averaging, the remaining leading finite-size error is not the generic energy-estimation problem. Instead it is an infrared density-correlation problem. This observation allows the correction to be extracted from the small-momentum static structure factor $S(\mbf q)$, rather than by repeating the full ground-state energy estimation routine on a sequence of larger simulation cells. We formulated the static-structure-factor operator in a Bloch-orbital basis, constructed its block encoding through the corresponding density operator, and showed how its ground-state expectation value can be estimated fault tolerantly using an amplified Hadamard test. We also introduced adaptive global and local procedures for identifying the infrared fitting window used to reconstruct the two-body finite size error contribution. The global procedure gives a direct and robust estimate of the stable small-momentum regime, while the local windowed procedure reduces the number of new structure-factor evaluations by reusing previously computed shell data. The resource analysis shows that this post-processing step remains subleading relative to the main ground-state energy estimation routine. The structure-factor correction has the same leading $(N_bN_k)^3$ dependence on the Bloch-orbital basis size that appears in representative periodic energy-estimation algorithms, but it avoids the large plane-wave prefactor associated with simulating the full Hamiltonian. Its logical-qubit overhead scales only as $\mathcal{O}(N_bN_k)$. 
\\

Thus, the algorithm targets a genuinely many-body observable while remaining below the cost of the dominant quantum primitive in the overall workflow.
The comparison with down-sampling highlights the practical importance of this separation. In the twist-averaged structure-factor approach, the energy-estimation cost is multiplied by the twist-averaging factor $\sqrt{M}$, and the two-body correction is then added through the subleading computation of $S(\mbf q)$. By contrast, down-sampling reaches the thermodynamic limit by repeatedly performing the most expensive part of the calculation on a hierarchy of larger Hamiltonians. Up to common precision factors, this produces a cost scaling of order $(2(N_aN_k)^{\frac{1}{3}}+1)(N_bN_k)^3N_{\rm pw}^{\frac{3}{2}}$, where the effective system size $N_aN_k$ would typically be much larger than the finite BvK cell used in the twist-averaged calculation. The difference is conceptual as much as asymptotic: our approach uses the quantum computer to measure the missing long-wavelength correlation directly, instead of inferring it from many repeated energy calculations as part of the classical post-processing step. This distinction is especially relevant for metals, conductors, and materials at extreme conditions. These systems often require dense $k$-space resolution to control shell effects and long-range screening, which can inflate quantum resource estimates well beyond the cost of a single finite-cell calculation. By combining twist averaging with a fault-tolerant static-structure-factor correction, the one-body and two-body pieces of the finite-size error are treated by observables naturally matched to their physical origin. This provides a route to smaller simulation cells, lower resource estimates, and more realistic quantum algorithms for extended materials.
\\

A few caveats are worth emphasizing. The number of twists $M$ required to make one-body shell effects negligible is not universal. It depends on the material, supercell geometry, electron density, metallic or insulating character, and the twist-sampling strategy, for example random twists, Monkhorst--Pack grids, special twists, or grand-canonical twist averaging. In practical calculations it may also depend on the many-body method used to estimate $E_{\btheta}$, such as QMC, diffusion Monte Carlo, coupled-cluster, or Full-CI, because method-dependent biases and statistical errors can vary with twist. Consequently, $M$ cannot generally be fixed a priori; it must be determined empirically from the convergence of the twist-averaged energy, or from a reliable proxy for its twist-to-twist variation. This makes fully material-specific quantum resource estimates for thermodynamic-limit ground-state energies more subtle than estimates for a single finite-cell calculation. There is substantial guidance on these issues from the QMC literature, including twist-averaged boundary conditions, special-twist constructions, grand-canonical twist averaging, and twist-selection strategies~\cite{Azadi_2019,Dagrada_2016_exact_special_twist,Mihm_2019_optimized_twist_ueg,Mihm_2021_twist_selection_beyond_mbpt,Mihm_2021_shortcut_metals,Annaberdiyev_2024}. An important direction for future work is to adapt these strategies to the ab initio Full-CI-like setting relevant for fault-tolerant quantum simulation. Doing so would enable a more complete and internally consistent resource-estimation framework for computing $E_{\infty}$, combining twist-averaged energy estimation with the SSF correction developed here.
\\

A further direction is to develop hybrid classical--quantum finite-size corrections in which the FTQC-SSF data is used to complement a classical continuum model. For example, one could compute a reference long-wavelength structure factor \(S_{\rm RPA}(q)\) from the RPA response, which captures the continuum screening behavior through the non-interacting Lindhard function, and use the quantum computer only to estimate the residual correlation correction
$
    \Delta S(q) = S_{\rm FTQC}(q)-S_{\rm RPA}(q).
$
continuous thermodynamic-limit surrogate could then be constructed as
$
    S_{\rm sur}(q)=S_{\rm RPA}(q)+\Delta S_{\rm fit}(q),
$
where \(\Delta S_{\rm fit}(q)\) is inferred from a small number of fault-tolerant measurements. Such a strategy could reduce the QRE associated to SSF evaluations while retaining sensitivity to correlation effects beyond RPA, and may be especially useful in systems where a simple infrared ansatz \(S(q)\sim \alpha q^\beta\) is insufficient, such as non-Fermi liquids, anisotropic metals, or materials with complicated multi-sheet Fermi surfaces. For more details and references on this see Appendix \ref{app: note on the IR ansatz}.
\\

Several other directions also remain open. First, the infrared reconstruction can be generalized beyond the leading form $S(q)\sim \alpha q^\beta$. A multi-term expansion $S(q)\sim \sum_i\sum_{ab} \alpha^{(i)}_{ab} (q_{a}q_{b})^{\beta_i}$ would allow subleading collective effects, anisotropy, and material-specific deviations from the leading asymptotic regime to be incorporated systematically \cite{Zaklama_2025}. Second, the static structure factor is important beyond finite-size corrections. It also appears in spectroscopy of crystalline solids \cite{Watanabe_1998,Mazzone_1983} and diffraction physics \cite{Svensson_1980,Zaklama_2025}, where density correlations determine how a material responds to probes such as x-rays, neutrons, or electrons. Extending the present algorithm to these settings could turn the same fault-tolerant primitive into a tool for computing response functions describing experimental data.

\section{Acknowledgements}
We would like to thank Stephan Eidenbenz, Brendan Krueger, Scott Pakin, Akram Touil and Yigit Subasi for many helpful discussions. J.G. and R.B. were supported by the Laboratory Directed Research and Development (LDRD) program of Los
Alamos National Laboratory (LANL) under project number 20260043DR as well as LANL’s ASC Beyond Moore’s
Law project. This research used resources provided by the Los Alamos National Laboratory Institutional
Computing Program. Los Alamos National Laboratory is operated by Triad National Security,
LLC, for the National Nuclear Security Administration of US Department of Energy (Contract No.
89233218CNA000001).
\appendix
\section{Shared regime for finite-size corrections}
\label{app: QMC_Bloch_equivalence}
Consider the three-dimensional homogeneous electron gas in a cubic periodic cell of side $L = N_aN_k$, and assume that the long-wavelength SSF obeys,
\begin{equation}
    S(q) = \alpha q+\mathcal{O}(q^{1+\nu}),
\end{equation}
where $\nu$ is positive, and non-zero.
Then the leading interaction finite-size correction obtained from the continuum-minus-mesh formula with the bare Coulomb kernel,
\begin{equation}
    \Delta V_{\text{Bloch}} = \frac{1}{2}\int \frac{d^3q}{(2\pi)^3}v(\mbf{q})\left(S(\mbf{q})-1\right) - \frac{1}{2\Omega}\sum_\mbf{q} v(\mbf{q})\left(S_{N_k}(\mbf{q})-1\right),
\end{equation}
which can be treated as a quadrature error.
Then the integral's argument becomes:
\begin{equation}
   d^3q v(q)S(q) = q^2 dq  \frac{1}{q} = q \ dq,
\end{equation}
which leads to a scaling of,
\begin{equation}
    \Delta V = \int_0^{q_{\text{min}}}  q dq = q^2_{\text{min}} = L^{-2}.
\end{equation}
If $v_\text{Ewald}=\frac{4\pi}{q^2}+\mathcal{O}(1)$ denotes the corresponding Ewald-periodized kernel, then the QMC-style correction, $\Delta V_{\text{QMC}}$, satisfies,
\begin{equation}
    \Delta V_{\text{Bloch}}-\Delta V_{\text{QMC}}= \mathcal{O}(L^{-4}),
\end{equation}
where the leading order correction between the two cancels as the leading order divergence of both Coulomb and Ewald potentials match exactly. Therefore,
\begin{equation}
    \Delta V_{\text{Bloch}} = \Delta V_{\text{QMC}} + o(L^{-2}).
\end{equation}
\\
Thus, the two formulations  for finite-size correction are asymptotically equivalent at leading order.
\section{The one-body finite-size error is sub-leading}
\label{app: DeltaU_is_subleading}

In the main text we argued that, once one-body shell effects have been reduced by sufficiently dense $k$-point sampling or by twist averaging, the dominant residual finite-size error is the two-body contribution encoded in $\Delta V$. The purpose of this appendix is to make that statement more explicit. We give two complementary arguments. The first is a direct reciprocal-space analysis of $\Delta U$, which makes the absence of infrared enhancement transparent. The second reformulates the twist-averaged one-body error as a BZ quadrature problem, yielding a more formal estimate of its decay with mesh density. Taken together, these arguments show that the one-body contribution is a genuine quadrature error, whereas the dominant two-body correction is controlled by long-wavelength density fluctuations.

\paragraph{Direct reciprocal-space argument.}
For a crystalline system, the one-body density may be written as
\begin{equation}
\rho(\mbf{q})=\sum_{\mbf{G}}\rho_{\mbf{G}}^{(\infty)}\delta(\mbf{q}-\mbf{G})~,
\qquad
\rho_{N_k}(\mbf{q})=\sum_{\mbf{G}}\rho_{\mbf{G}}^{(N_k)}\delta_{\mbf{q},\mbf{G}}~,
\end{equation}
so that the infinite- and finite-mesh one-body contributions reduce to reciprocal-lattice sums,
\begin{equation}
U_{\infty}
=
\int_{\rm BZ} d^3\mbf{q}\,u(\mbf{q})\rho(\mbf{q})
=
\sum_{\mbf{G}\in{\rm BZ}}\rho_{\mbf{G}}^{(\infty)}u(\mbf{G})~,
\qquad
U_{N_k}
=
\sum_{\mbf{G}\in{\rm BZ}}\rho_{\mbf{G}}^{(N_k)}u(\mbf{G})~.
\end{equation}
Their difference is therefore
\begin{equation}
\Delta U
=
\sum_{\mbf{G}\in{\rm BZ}}
\Big(\rho_{\mbf{G}}^{(\infty)}-\rho_{\mbf{G}}^{(N_k)}\Big)u(\mbf{G})~.
\end{equation}
The key observation is that the potentially dangerous long-wavelength contribution is absent: charge neutrality implies $u(0)=0$, so the $\mbf{q}=0$ mode does not contribute. Thus, unlike the two-body term, $\Delta U$ does not receive infrared enhancement from long-range Coulomb fluctuations. To estimate the remaining error, write the reciprocal-space coefficients as the BZ average and the corresponding uniform $k$-mesh average of a smooth periodic function $g_{\mbf G}(\mbf k)$,
\begin{align}
\rho_{\mbf{G}}^{(\infty)} &= \frac{1}{\Omega_{\rm BZ}}\int_{\rm BZ} d^3\mbf{k}\; g_{\mbf{G}}(\mbf{k})~,\\
\rho_{\mbf{G}}^{(N_k)} &= \frac{1}{N_k}\sum_{\mbf{k}\in{\rm BZ}} g_{\mbf{G}}(\mbf{k})~.
\end{align}
Expanding $g_{\mbf G}$ in Fourier modes,
\begin{equation}
g_{\mbf{G}}(\mbf{k})
=
\sum_{\mbf{m}\in\mathbb{Z}^3}
e^{i\mbf{m}\cdot \mbf{k}}\,
\tilde g_{\mbf{G}}(\mbf{m})~,
\end{equation}
the continuum average retains only the zero mode, whereas the discrete mesh samples only modes compatible with the mesh periodicity. If $M\sim N_k^{1/3}$ is the linear mesh size, then
\begin{equation}
\rho_{\mbf{G}}^{(N_k)}
=
\sum_{\boldsymbol{\ell}\in\mathbb{Z}^3}\tilde g_{\mbf{G}}(M\boldsymbol{\ell})~,
\qquad
\Delta\rho_{\mbf{G}}
:=
\rho_{\mbf{G}}^{(\infty)}-\rho_{\mbf{G}}^{(N_k)}
=
-\sum_{\boldsymbol{\ell}\in\mathbb{Z}^3\setminus\{\mbf{0}\}}\tilde g_{\mbf{G}}(M\boldsymbol{\ell})~.
\end{equation}
Assume the Fourier coefficients decay algebraically,
\begin{equation}
|\tilde g_{\mbf{G}}(\boldsymbol{\ell})|
\le \frac{C_{\mbf{G}}}{|\boldsymbol{\ell}|^{p}},
\qquad p>0~.
\end{equation}
This is a conservative assumption: many insulators, and even some metals away from singular regimes, exhibit faster---often exponential---decay. Then
\begin{equation}
|\Delta\rho_{\mbf{G}}|
\le
\frac{C_{\mbf{G}}}{M^{p}}
\sum_{\boldsymbol{\ell}\in\mathbb{Z}^3\setminus\{\mbf{0}\}}
\frac{1}{|\boldsymbol{\ell}|^{p}}~.
\end{equation}
In the ideal case, this upper bound should remain finite. It represents the limiting situation in which all coefficients $\tilde g_{\mbf G}$ are positive and decay in the same manner. Since such behavior cannot, in general, be excluded for a given material, finiteness of the sum is an important consistency requirement. The lattice sum converges for $p>3$, in which case
\begin{equation}
\Delta\rho_{\mbf{G}}
=
\mathcal{O}(M^{-p})
=
\mathcal{O}\!\left(N_k^{-p/3}\right)~.
\end{equation}
This is already sufficient to show that $\Delta U$ decays faster than an $\mathcal{O}(N_k^{-1})$ infrared correction whenever $p>3$. More importantly, the physical content is clear irrespective of the precise exponent: $\Delta U$ is a BZ quadrature error, not a long-wavelength collective effect. This contrasts sharply with the two-body contribution. There the relevant quantity is the density fluctuation governed by $\rho(\mbf{q})$, so the long-wavelength sector enters directly through the SSF. For Coulomb interactions, $v(\mbf q)\sim 1/q^2$, which strongly weights the smallest nonzero $\mbf q$ modes. Consequently, once the one-body shell error has been reduced, the dominant residual finite-size effect is controlled by the missing small-$q$ region and therefore by $\Delta V$, not by $\Delta U$.

\paragraph{Twist-averaged quadrature estimate.}
The same conclusion can be reached in a more formal way by considering the one-body error after twist averaging. Write the total finite-size error schematically as
\begin{equation}
    \Delta E_{\rm FS}
    =
    \Delta U(n)
    \;+\;
    \Delta V(n)
    \;+\; \cdots~,
\end{equation}
where $n=N_k^{1/3}$ is the linear twist-grid size, $\Delta U$ is the residual one-body error after twist averaging, and $\Delta V$ is the two-body Coulomb correction captured by $\Delta V$. The first term is simply the error in approximating the BZ average by a uniform grid,
\begin{equation}
    \Delta U
    =
    |\hat{E}_n-E_\infty|~.
\end{equation}

Let $E(\theta)$ denote the twist-dependent energy, where $\theta$ labels boundary-condition phases on the three-dimensional torus $\mathcal{T}^3$. Twist averaging is then a periodic quadrature rule on $\mathcal{T}^3$, and its convergence is governed by the smoothness of $E(\theta)$. If $E(\theta)\in W^{s,\infty}(\mathcal{T}^3)$, the Sobolev space of functions on the 3D torus whose derivatives up to order $s$ are bounded, periodic quadrature theory gives \cite{Trefethen_2014,Kazashi_2023}
\begin{equation}
    \Delta U
    \leq
    C_s n^{-s}
    =
    C_s N_k^{-s/3}~.
\end{equation}
If $E(\theta)$ is analytic, the convergence improves to an exponential form,
\begin{equation}
    \Delta U
    \leq
    C_s e^{-\alpha n}
    =
    C_s e^{-\alpha N_k^{1/3}}~,
\end{equation}
as expected for analytic periodic quadrature \cite{Dick_2017}.
\\

This estimate makes the role of twist averaging particularly transparent. For sufficiently smooth twist dependence, the one-body shell error decreases rapidly with $N_k$, becoming parametrically smaller than the leading two-body correction. In metallic systems, where Fermi-surface crossings reduce smoothness, this convergence is weakened and may revert to lower-order algebraic behavior. In that regime one may write more generally
\begin{equation}
    \Delta U
    =
    \mathcal{O}(N_k^{-p})~,
\end{equation}
with an exponent $p$ determined by the regularity of the twist-dependent energy and by any smoothing introduced through finite-temperature occupations or smearing. Even then, the one-body term remains a quadrature error rather than an infrared Coulomb singularity. By contrast, Appendix~\ref{app: QMC_Bloch_equivalence} shows that the leading two-body correction behaves as
\begin{equation}
    \Delta V
    =
    \mathcal{O}(L^{-1})
    \quad \text{or} \quad
    \mathcal{O}(L^{-2})~.
\end{equation}
Thus, once the twist grid is sufficiently dense and the shell error is brought under control, one finds
\begin{equation}
    \Delta U \ll \Delta V~.
\end{equation}

The two arguments above address the same point from complementary directions. The reciprocal-space analysis shows physically why $\Delta U$ is sub-leading: the $\mbf q=0$ contribution cancels by neutrality, so the one-body term does not inherit the infrared enhancement that governs the two-body correction. The twist-averaged quadrature analysis then makes the convergence rate explicit, showing that the remaining one-body error decays with the smoothness of the twist-dependent energy. Together they justify the approximation used throughout the main text: after shell effects have been reduced, the leading residual finite-size error is controlled by the two-body correction $\Delta V$.

\section{Amplitude amplification for the Hadamard-test}
\label{app: amplitude_amplification}

This appendix summarizes the amplitude-amplification construction used in Section~\ref{sec: quantum_algorithm}. Its role is to estimate the Hadamard-test success probability $P(0)$ more efficiently than by direct sampling, thereby reducing the cost of estimating $S^{(\btheta)}_{N_k}(\mbf q)$.
\\

Starting from the Hadamard-test state
\begin{equation}
    |\varphi\rangle
    =
    U_{\varphi}\ket{0}\ket{0^m}\ket{0^k},
\end{equation}
we decompose it as
\begin{equation}
    |\varphi\rangle
    =
    \sqrt{P(0)}\,|\psi_{\rm good}\rangle
    +
    \sqrt{1-P(0)}\,|\psi_{\rm bad}\rangle~,
\end{equation}
where the good subspace is defined by the control qubit being in the state $\ket{0}$. The corresponding projector is
\begin{equation}
    \Pi_{\rm good}=|0\rangle\langle 0|\otimes I_{k+m}~.
\end{equation}
The standard Grover iterate is then
\begin{equation}
    G = R_{\varphi}R_{\rm good}~,
\end{equation}
where $R_{\rm good}$ flips the phase of the good subspace and $R_{\varphi}$ reflects about the prepared state $|\varphi\rangle$. In circuit form,
\begin{center}
\begin{quantikz}
\lstick{$\ket{0}$}      & \gate[wires=3]{G} & \qw \\
\lstick{$\ket{0^{m}}$}  &  & \qw\\
\lstick{$\ket{0^{k}}$}  &  & \qw
\end{quantikz}
$\;=\;$
\begin{quantikz}
\lstick{$\ket{0}$} & \gate[wires=2]{R_{\rm good}} & \gate[wires=2]{R_{\varphi}} & \qw \\
\lstick{$\ket{0^{k+m}}$} & & & \qw
\end{quantikz}
\end{center}
The reflection about $|\varphi\rangle$ is implemented by unpreparing the Hadamard-test state, reflecting about the all-zero computational basis state, and preparing $|\varphi\rangle$ again:
\begin{center}
\begin{quantikz}
\lstick{$\ket{0}$}      & \gate[wires=2]{R_{\varphi}} & \qw \\
\lstick{$\ket{0^{k+m}}$} &  & \qw
\end{quantikz}
$\;=\;$
\begin{quantikz}
\lstick{$\ket{0}$} & \gate[wires=2]{U_{\varphi}^{\dg}} & \gate[wires=2]{\mathrm{REF}} & \gate[wires=2]{U_{\varphi}} & \qw \\
\lstick{$\ket{0^{k+m}}$} & & & & \qw
\end{quantikz}
\end{center}
where
\begin{equation}
    \mathrm{REF}=2\ket{0^{k+m+1}}\bra{0^{k+m+1}}-I~.
\end{equation}
The good-space reflection is especially simple, since the good subspace is identified entirely by the control qubit:
\begin{center}
\begin{quantikz}
\lstick{$\ket{0}$}      & \gate[wires=2]{R_{\rm good}} & \qw \\
\lstick{$\ket{0^{k+m}}$} &  & \qw
\end{quantikz}
$\;=\;$
\begin{quantikz}
\lstick{$\ket{0}$} & \gate{-Z} & \qw \\
\lstick{$\ket{0^{k+m}}$} & \qw & \qw
\end{quantikz}
\end{center}
Equivalently,
\begin{align}
    R_{\varphi}
    &= U_{\varphi}
    \bigl(2\ket{0^{k+m+1}}\bra{0^{k+m+1}}-I\bigr)
    U_{\varphi}^{\dg},\\
    R_{\rm good}
    &= I-2\Pi_{\rm good}
    = -Z\otimes I_{k+m},
\end{align}
up to an irrelevant global phase convention. As usual, $G$ acts as a rotation in the two-dimensional subspace spanned by $|\psi_{\rm good}\rangle$ and $|\psi_{\rm bad}\rangle$. Its nontrivial eigenvalues are $e^{\pm 2i\theta}$, where
\begin{equation}
    \sin^2\theta = P(0)~.
\end{equation}
Thus, estimating $P(0)$ is equivalent to estimating the phase $\theta$. This is done by applying QPE to the Grover iterate. The corresponding circuit is
\begin{center}
\begin{quantikz}
\lstick{$\ket{0^d}$}      & \gate{H^{\otimes d}} & \gate[wires=4]{\mathcal{G}} & \gate{U^{\dg}_{\rm QFT}} & \meter{} \\
\lstick{$\ket{0}$}        & \qw &  & \qw & \qw \\
\lstick{$\ket{0^{k+m}}$}  & \qw &  & \qw & \qw \\
\lstick{$\ket{\psi_{+}}$} & \qw &  & \qw & \qw
\end{quantikz}
\end{center}
where
\begin{equation}
    \mathcal{G}=\sum_{j\in [2^d]}|j\rangle\langle j|\,G^j,
\end{equation}
and
\begin{equation}
    \ket{\psi_{+}}
    =
    \frac{1}{\sqrt{2}}
    \left(
    |\psi_{\rm good}\rangle
    +
    i\,|\psi_{\rm bad}\rangle
    \right).
\end{equation}
The phase register has size $d=\log \epsilon_P^{-1}$, and measuring it yields an estimate of $\theta$, hence of $P(0)$.
\\

The key consequence is the standard amplitude-estimation scaling: the success probability $P(0)$, and therefore $S^{(\btheta)}_{N_k}(\mbf q)$, can be estimated with cost $\mathcal{O}(\epsilon_P^{-1})$ rather than the $\mathcal{O}(\epsilon_P^{-2})$ repetitions required by direct sampling. If the prepared state has overlap $p_0$ with the true ground state, the usual overlap penalty appears, giving the overall query complexity
\begin{equation}
    \mathcal{O}\!\left(p_0^{-1/2}\,\epsilon_P^{-1}\right).
\end{equation}
This is the scaling used in the resource estimates of Section~\ref{sec: quantum_algorithm}.
\\

Finally, the cost of a single Grover iterate is set by the cost of the Hadamard-test preparation $U_\varphi$, its inverse, and the reflection about the all-zero state. Since $U_\varphi$ is built from $U_\psi$ and $U_{\hat S}$, the per-query logical cost is
\begin{equation}
    T_{U_\psi}
    +T_{U_{\hat S}}
    +T_{U_\psi^\dg}
    +T_{U_{\hat S}^\dg}
    +T_{\rm REF},
\end{equation}
as quoted in the main text.
\section{Derivation of the noise-tolerance bound for \texorpdfstring{$S(\mbf q)$}{S(q)}}
\label{app:noise_tolerance_details}

In this appendix we derive the noise-tolerance condition quoted in Section~\ref{sec:noise_tolerance_main} and the resulting asymptotic cost of estimating the leading finite-size correction $\Delta V$. Throughout, we make the simplifying assumption that the dominant residual finite-size error is carried by the two-body correction $\Delta V$, so that subleading terms may be neglected at the level of the present error budget.
\\

We begin by partitioning the total target error $\epsilon$ equally between the raw finite-system energy density estimate and the correction $\Delta V$:
\begin{equation}
    \epsilon^2 = \sigma_{E}^2+\sigma_{\Delta V}^2
    = \frac{\epsilon^2}{2}+\frac{\epsilon^2}{2}~.
\end{equation}
We then split the uncertainty in $\Delta V$ equally between the discrete reciprocal-space sum and the continuum integral,
\begin{equation}
    \sigma_{\rm sum}^2=\frac{\sigma_{\Delta V}^2}{2},
    \qquad
    \sigma_{\rm integral}^2=\frac{\sigma_{\Delta V}^2}{2}~.
\end{equation}
Both contributions are controlled by the noise in the computed values of $S(\mbf q)$. To determine how this noise propagates into the correction, we fit the infrared form $S(q)\sim \alpha q^\beta$ using sampled data $\{|\mbf{q}_i|\}_{i=1}^{n_{\rm sample}}$. The statistical uncertainty in the fitted slope is
\begin{equation}
    \sigma_{\alpha}
    =
    \frac{\sigma_{S_{N_k}^{(\btheta)}}}
    {\sqrt{\sum_{\btheta}\sum_{i=1}^{f(n_{\rm sample})} |\mbf{q}_i|^{2\beta}}}~,
    \label{eq:error_alpha}
\end{equation}
where $\sigma_{S_{N_k}^{(\btheta)}}$ is the simulation error associated with the evaluation of $S(q)$, taken to be independent of $q$, and $f(n_{\rm sample})<n_{\rm sample}$ accounts for the fact that only a subset of the sampled points may lie in the clean asymptotic regime. The induced uncertainty in the continuum part of $\Delta V$ is then
\begin{equation}
    \sigma_{\rm integral}=C(N_k)\sigma_\alpha~,
    \label{eq:error_integral}
\end{equation}
with
\begin{equation}
    C(N_k)=\frac{1}{4\pi^2}\int_{q\le q_t} dq\, q^{\beta+2}v(q)~.
\end{equation}
We now impose a concrete noise model by writing
\begin{equation}
    \sigma_{S_{N_k}^{(\btheta)}}=p\;a\,\,\epsilon~,
    \qquad
    a=(\Omega_{\rm unit})^{1/3}~,
\end{equation}
where $p>0$ is a dimensionless noise prefactor and the factor $a$ is introduced for dimensional consistency. For the Coulomb interaction $V(|\mbf r'-\mbf r|)=1/|\mbf r'-\mbf r|$, one has $ v(q)=\frac{4\pi}{q^2}$ so that after doing the integral we get
\begin{equation}
    C(N_k)
    =
    \frac{q_t^{\beta+1}}{\pi(\beta+1)}~.
\end{equation}
Equation~\eqref{eq:error_integral} therefore gives
\begin{equation}
    \sigma_{\rm integral}
    =
    \frac{\epsilon}{2}
    =
    \frac{q_t^{\beta+1}}{\pi(\beta+1)}
    \frac{pa\epsilon}{\sqrt{\sum_{\btheta}\sum_{i=1}^{f(n_{\rm sample})} |\mbf{q}_i|^{2\beta}}}~.
\end{equation}
While the error coming from the discrete sum is
\begin{equation}
    \sigma_{\rm sum} = \frac{\epsilon}{2}
    =
    \frac{ p\epsilon}{2(N_aN_kM)a^2}
    \sqrt{
        \sum_{\btheta}\sum_{\mbf q;\,|\mbf q|\le q_t}
        \frac{1}{|\mbf q|^4}
    }~.
\end{equation}
Our goal is therefore to convert the two requirements
$\sigma_{\rm integral}=\epsilon/2$ and $\sigma_{\rm sum}=\epsilon/2$
into explicit upper bounds on $p$. To do so, we parameterize the infrared shell vectors as
\begin{equation}
    \mbf q=\frac{2\pi}{a(N_aN_k)^{1/3}}\mbf m~,
\end{equation}
with the truncation vector being such that $|\mbf m_t|=\lceil \eta_k^{\frac{1}{3}}\rceil ~.$
The integer vectors $\mbf m_i$ have components in
$\{0,1,\dots,\lceil \eta_k^{\frac{1}{3}}\rceil\}$, but we keep only distinct magnitudes. Accordingly, we define
\begin{equation}
    \mathcal S
    =
    \left\{
    \mbf m
    \;\middle|\;
    |\mbf m'|\neq |\mbf m|
    \ \forall\,\mbf m'\neq \mbf m \in \mathbb{Z}^3,
    \ \text{and}\ 
    |\mbf m|\le \lceil \eta_k^{\frac{1}{3}}\rceil
    \right\},
\end{equation}
so that
\begin{equation}
    \sum_{i=1}^{f(n_{\rm sample})} |\mbf{q}_i|^{2\beta}
    =
    \left(\frac{2\pi}{a(N_aN_k)^{1/3}}\right)^{2\beta}
    \sum_{\mbf m\in\mathcal S}(\mbf m)^{2\beta}~.
\end{equation}
Using the bound
\begin{equation}
    \sum_{\mbf m\in\mathcal S}(\mbf m)^{2\beta}
    \le
    \frac{(\eta_k^{2/3}+1)^{\beta+1}-1}{\beta+1}~,
\end{equation}
which comes from using the sum-integral inequality $\sum_{x=0}^{\mu}f(x) \leq \int_{1}^{\mu+1}f(x)\;dx$ for $f(x)$ being a strictly increasing integrable function. This when combined with
\begin{equation}
    C(N_k)
    =
    \left(\frac{2\pi}{a(N_aN_k)^{1/3}}\right)^{\beta+1}
    \frac{(\eta_k^{\frac{1}{3}})^{\beta+1}}{\pi(\beta+1)}~,
\end{equation}
one finds the first constraint on $p$ for large $\eta_k$:
\begin{align}
    p
    \le
\frac{1}{4}\sqrt{M(\beta+1)}(N_aN_k)^{1/3}\left(1+\frac{\beta+1}{2\eta_k^{\frac{2}{3}}}\right)~.
\end{align}
A second constraint follows from the discrete-sum error. Summing now over all $\mbf m^{(\btheta)}$ with $|\mbf m^{(\btheta)}|\le \eta_k^{\frac{1}{3}}$, we obtain
\begin{equation}
    \sum_{\mbf q;\,|\mbf q|\le q_t}
        \frac{1}{|\mbf q|^4}
    =
    \left(\frac{a(N_aN_k)^{1/3}}{2\pi}\right)^4
    \sum_{\mbf m\in \mathbb{Z}^3;\,|\mbf m|\neq 0}^{\eta_k}\frac{1}{(\mbf m)^4}~.
\end{equation}
Using the estimate
\begin{equation}
    \sum_{\mbf m\in \mathbb{Z}^3;\,|\mbf m|\neq 0}^{\eta_k}\frac{1}{(\mbf m)^4}
    \le
    S_{\infty}-\frac{\pi}{2\eta_k^{\frac{1}{3}}}~,
\end{equation}
with $S_\infty\simeq 16.53$, one finds
\begin{equation}
   p \leq \frac{(N_kN_a)^{1/3}\sqrt{M}}{2\sqrt{S_{\infty}}}\left(1+\frac{\pi}{4\eta_k^{\frac{1}{3}}S_{\infty}}\right)~.
\end{equation}
Combining the integral and discrete-sum constraints gives the admissible noise bound
\begin{equation*}
    p \leq \mathfrak{p}\sqrt{M}N_k^{\frac{1}{3}}~.
\end{equation*}
In the main text, however, the factor of $\sqrt{M}$ is absorbed into the overall query-complexity prefactor. Specifically, the total number of queries to the block encoding of $\hat{S}$ contributes a prefactor $M/\epsilon$, so absorbing the $\sqrt{M}$ from the noise bound leaves an overall factor $\sqrt{M}/\epsilon$ multiplying the Toffoli cost of a single query to the oracle that computes $S^{(\btheta)}_{\mbf{q}+\btheta}$. Therefore, in the main text we quote the normalized bound
\begin{equation}
    p \leq \mathfrak{p}N_k^{\frac{1}{3}}~,
    \label{eq:p_bound_appendix}
\end{equation}
where
\begin{equation}
    \mathfrak{p}
    =
    \frac{N_a^{\frac{1}{3}}}{2}\left(\min\left\{
        \frac{1}{\sqrt{S_{\infty}}}
        \left(1+\frac{\pi}{4\eta_k^{\frac{1}{3}}S_{\infty}}\right),
        \frac{1}{2}\sqrt{\beta+1}
        \left(1+\frac{\beta+1}{2\eta_k^{\frac{2}{3}}}\right)
    \right\}\right)~.
\end{equation}
This is the origin of Eq.~\eqref{eq:p_bound} in the main text.
\section{Derivation of the LCU form of the density operator}
\label{app:lcu_density_details}

In this appendix we derive the linear-combination-of-unitaries representation of the density operator $\hat\rho_{\mbf q}$ used in Section~\ref{sec:lcu_density_main}, together with its associated one-norm and the resulting scaling of the block encoding of $\hat S_{\mbf q}$.
\\

We begin from the observation that the SSF can be block encoded once a block encoding of the density operator is available. Specifically,
$
    U_{\hat{S}}
    =
    [U_{\hat{\rho}}]^{\dg}
    \cdot
    \mathrm{REF}
    \cdot
    U_{\hat{\rho}}~,
$
which implements a block encoding of
$
    \frac{\hat{S}_{\mbf{q}}}{\tfrac{1}{2}\left(\lambda_{\rho}\right)^2}-1~.
$
It follows that the normalization factor for $\hat S_{\mbf q}$ is
\begin{equation}
    \lambda_{\hat{S}} = \frac{1}{2}\left(\lambda_{\rho}\right)^2~.
\end{equation}
The main task is therefore to obtain an LCU representation of $\hat\rho_{\mbf q}$ and determine its one-norm $\lambda_\rho$. For convenience of notation, in this calculation we suppress the twist label by absorbing it into the lattice vector $\mbf{q}$, and then reintroduce it in the final equation. Our starting point is the form-factor tensor in a fixed momentum-transfer sector,
\begin{equation}
    \mbf{F}_{\mbf{k}}(\mbf{q})
    =
    \sum_{i,j}
    F_{\mbf{k}i,\mbf{k}\oplus\mbf{q} j}
    \hat{E}_{\mbf{k}i,\mbf{k}\oplus\mbf{q}j}~.
\end{equation}
Since $\hat\rho_{\mbf q}$ is not Hermitian in general, it is convenient to separate it into two Hermitian pieces. We therefore write
\begin{equation}
    \mbf{F}
    =
    \boldsymbol{\eta}^{(0)} \;+\; i\,\boldsymbol{\eta}^{(1)}~,
\end{equation}
where
\begin{equation}
    \boldsymbol{\eta}^{(J)}
    \equiv
    \frac{\mbf{F}\;+\;(-1)^{J}\mbf{F}^{\dg}}{2i^{J}},
    \qquad
    J \in \{0,1\}~.
\end{equation}
By construction, each $\boldsymbol{\eta}^{(J)}$ is Hermitian. To express these operators in a diagonalizable single-particle form, we introduce the vector of annihilation operators
\begin{equation}
\mbf a_{\mbf{k},\sigma}(\mbf{q})
=
\big(
\hat a_{\mbf{k},1,\sigma},\ldots,\hat a_{\mbf{k},N_b,\sigma},
\hat a_{\mbf{k}\oplus\mbf{q},1,\sigma},\ldots,\hat a_{\mbf{k}\oplus\mbf{q},N_b,\sigma}
\big)^{\mathsf T}.
\end{equation}
In this notation, $\boldsymbol{\eta}^{(J)}_{\mbf k}(\mbf q)$ can be written as
\begin{equation}
\boldsymbol{\eta}^{(J)}_{\mbf{k}}(\mbf{q})
=
\frac{1}{2i^{\delta_{J 1}}}
\sum_{\sigma}\sum_{\mbf{k}}
\mbf a_{\mbf{k},\sigma}^{\dg}\;
\begin{pmatrix}
0 & F_{\mbf{k}}(\mbf{q})\\
(-1)^{J}F_{\mbf{k}}^{\dg}(\mbf{q}) & 0
\end{pmatrix}\;
\mbf a_{\mbf{k},\sigma}~,
\label{eq:fourier_matrix_decompisition_appendix}
\end{equation}
where,
\begin{equation}
    (F_{\mbf{k}}(\mbf{q}))_{ij} := F_{\mbf{k}i,\mbf{k}\oplus\mbf{q}j}~.
\end{equation}
The matrix in Eq.~\eqref{eq:fourier_matrix_decompisition_appendix} is Hermitian, and therefore admits a unitary diagonalization:
\begin{equation}
    \hat{\mbf{U}}^{(J)\dg}_{\mbf{k}}(\mbf{q})\;
\Bigg[
\frac{1}{2 i^{\delta_{J 1}}}
\begin{pmatrix}
0 & F_{\mbf{k}}(\mbf{q})\\
(-1)^{J}F_{\mbf{k}}^{\dg}(\mbf{q}) & 0
\end{pmatrix}
\Bigg]
\hat{\mbf{U}}^{(J)}_{\mbf{k}}(\mbf{q})
=
\operatorname{diag}\!\big(f^{(J)}_{1},\ldots,f^{(J)}_{N_b}\big)~.
\end{equation}
Substituting this diagonal form back into $\boldsymbol{\eta}^{(J)}$ gives
\begin{equation}
    \boldsymbol{\eta}^{(J)}
    =
    \frac{1}{2i^{\delta_{J 1}}}
    \sum_{\sigma}\sum_{\mbf{k}}
    \mbf a_{\mbf{k},\sigma}^{\dg}\;
    \hat{\mbf{U}}^{(J)}_{\mbf{k}}(\mbf{q})\;
    \operatorname{diag}\!\big(f^{(J)}_{1},\ldots,f^{(J)}_{N_b}\big)\;
    \hat{\mbf{U}}^{(J)\dg}_{\mbf{k}}(\mbf{q})\;
    \mbf a_{\mbf{k},\sigma}~.
\end{equation}
We now define the rotated annihilation operators
\begin{equation}
\mbf b_{\mbf{k},\sigma}^{(J)}
=
\hat{\mbf{U}}^{(J)\dg}(\mbf{q})\;
\mbf a_{\mbf{k},\sigma}\;
\hat{\mbf{U}}^{(J)}(\mbf{q}),
\qquad
\hat n^{(J)}_{i,\mbf{k},\sigma}
=
\hat b_{i,\mbf{k},\sigma}^{(J)\dg}\;
\hat b_{i,\mbf{k},\sigma}^{(J)}~.
\end{equation}
In terms of these operators, the diagonalized form becomes
\begin{align}
    \boldsymbol{\eta}^{(J)}
    &=
    \sum_{\mbf{k}}^{N_k}\sum_{i=1}^{N_b/2}\sum_{\sigma}
    \hat{\mbf{U}}^{(J)}_{\mbf{k}}(\mbf{q})\;
    \hat b^{(J)\dg}_{\mbf{k},i,\sigma}\;
    f^{(J)}_{i}(\mbf{q},\mbf{k})\;
    \hat b^{(J)}_{\mbf{k},i,\sigma}\;
    \hat{\mbf{U}}^{(J)\dg}_{\mbf{k}}(\mbf{q}) \nonumber\\
    &=
    \sum_{\mbf{k}}^{N_k}\sum_{i=1}^{N_b/2}\sum_{\sigma}
    \sign(f^{(J)}_{i})|f^{(J)}_{i}(\mbf{q},\mbf{k})|
    \;\hat{\mbf{U}}^{(J)}_{\mbf{k}}(\mbf{q})\;
    \hat n^{(J)}_{\mbf{k},i,\sigma}\;
    \hat{\mbf{U}}^{(J)\dg}_{\mbf{k}}(\mbf{q})~.
\end{align}
Using the identity $\hat n_\alpha=(I-Z_\alpha)/2$, and absorbing the factor of $-i^J$ into the definition of the unitaries, then we obtain the LCU decomposition
\begin{equation}
    \hat{\rho}_{\mbf{q}}
    =
    \sum_{J=0,1} i^{J}\boldsymbol{\eta}^{(J)}
    =
    \frac{1}{2}\sum_{J=0,1}\sum_{\mbf{k}}^{N_k}\sum_{i=1}^{N_b/2}\sum_{\sigma}\sign(f^{(J)}_{i})|
    f^{(J)}_{i}(\mbf{q},\mbf{k})|\;
    \hat{\mbf{U}}^{(J)}_{\mbf{k}}(\mbf{q})\;
    Z^{(J)}_{\mbf{k},i,\sigma}\;
    \hat{\mbf{U}}^{(J)\dg}_{\mbf{k}}(\mbf{q})~,
\end{equation}
where we have discarded the identity term as it results in a sub-leading contribution to the asymptotic cost in the basis size.
The corresponding one-norm is
\begin{equation}
    \lambda_{\rho}
    =
    \sum_{J=0,1}\sum_{\mbf{k}}^{N_k}\sum_{i=1}^{N_b/2}
    \left|f^{(J)}_{i}(\mbf{q},\mbf{k})\right|~.
\end{equation}
This form is particularly useful because it isolates the entire dependence on the material and on the momentum transfer $\mbf q$ into the coefficients $f_i^{(J)}(\mbf q,\mbf k)$ and the corresponding Givens rotations $\hat{\mbf U}^{(J)}_{\mbf k}(\mbf q)$. Once these are known, the many-body operator is represented as a weighted sum of conjugated Pauli $Z$ terms, which is precisely the structure needed for LCU block encoding.
\section{Detailed resource estimates for block encoding the density operator}
\label{app:block_encoding_density_details}

In this appendix we collect the explicit block-encoding constructions and resource estimates for the density operator $\hat\rho_{\mbf q}$. We begin with the $\Gamma$-point specialization, where the lattice-momentum sector structure collapses to a single sector, and then turn to the general case away from the $\Gamma$-point. The latter is the construction used in the main text, but the $\Gamma$-point case is a useful warm-up and makes the organization of the $\PREP$ and $\SEL$ stages more transparent.

\subsection{At the \texorpdfstring{$\Gamma$}{Gamma}-point}

At the $\Gamma$-point we consider only the mode $\mbf Q=\mbf k=0$, so the density operator in the LCU representation reduces to
\begin{equation}
\hat{\rho}_{\mbf{q}}\big|_{\Gamma} \equiv \hat{\rho} = \frac{1}{2}\sum_{J=0,1}\sum_{i=1}^{N_b/2}\sign(f^{(J)}_i)|f^{(J)}_i|\hat{\mbf{U}}^{(J)}\;Z^{(J)}_i\;\hat{\mbf{U}}^{(J)\dg}~.
\end{equation}
Our goal is to encode this operator as a block of a larger unitary matrix. To this end, we define the walk operator
\begin{equation}
    \hat{\mathcal{Q}}_{W} = \PREP\cdot\SEL\cdot\PREP^{\dg}~,
\end{equation}
following the prepare--select framework of Ref.~\cite{Lin_notes}. This construction yields the desired block encoding $U_{\hat\rho}$. We adopt the QROAM-based linear-$T$ prescription of Ref.~\cite{Babbush_2018}, specialized to a general one-body Hermitian operator of the form
\begin{equation}
    \hat{O} = \sum_{p,q}^{N_b/2}\sum_{\sigma}o_{pq}\;\hat{a}_{p,\sigma}^{\dg}\hat{a}_{q,\sigma}~.
    \label{eq:O_operator_appendix}
\end{equation}
We begin with the $\PREP$ operation, whose purpose is to prepare
\begin{equation}
    \PREP\cdot|0^{n}\rangle|0\rangle|0\rangle = |L\rangle = \frac{1}{\sqrt{2\lambda_{O}}}\sum_{\sigma}\sum_{p=1}^{N_b/2}\sqrt{|\mu_p|}\;|p\rangle\;|\sigma\rangle\;|s_p\rangle~,
\end{equation}
where $\lambda_O$ is the one-norm of the operator in Eq.~\eqref{eq:O_operator_appendix}, $n=\log (N_b/2)$, $\mu_p$ are the eigenvalues in the spatial-orbital basis, and $s_p\in\{0,1\}$ stores their sign. The first stage prepares the spin qubit and an equal superposition over orbital labels, with resource cost
\begin{align}
    \text{Toffoli} &= 3n-3\eta+2b_r-9~,\\
    \text{Qubit} &= n+1+1~,
\end{align}
where $\eta$ is the highest power of $2$ dividing $|\mu_p|$ and $b_r$ is the bit precision of the rotation angles. The final step of $\PREP$ loads the sign and magnitude data using QROAM-based coherent alias sampling~\cite{Babbush_2018}. This contributes
\begin{equation}
    \text{Toffoli}
    =
    \left\lceil\frac{N_b}{2k_P}\right\rceil
    +(n+\mathfrak N+1)(k_P-1)
    +\mathfrak N+n~,
\end{equation}
and
\begin{equation}
    \text{Qubit}
    =
    n+\mathfrak N+1+1+1+1~,
\end{equation}
where $k_P$ is the QROAM parameter and $\mathfrak N$ is the alias-sampling parameter. Collecting the two contributions gives
\begin{equation}
    T_{\PREP}
    =
    \left\lceil\frac{N_b}{2k_P}\right\rceil
    +(n+\mathfrak N+1)(k_P-1)
    +\mathfrak N+4n-3\eta+2b_r-9~,
\end{equation}
with peak qubit cost
\begin{equation}
    Q_{\PREP}=2n+\mathfrak N+6~.
\end{equation}
Next we construct the $\SEL$ operation. Conditioned on the state $|p\rangle|\sigma\rangle|s_p\rangle$, it applies the transformation
\begin{equation}
    (p,\sigma)\mapsto (-1)^{s_p}\hat U_p\,Z_{p,\sigma}\hat U_p^\dg~.
\end{equation}
This requires QROAM loading of the Givens-rotation data, spin-controlled swaps, implementation of $\hat U_p$, the selected $Z$ rotation, and subsequent uncomputation. The individual contributions combine to
\begin{align}
    T_{\SEL}
    &=
    \left\lceil\frac{N_b}{2k_r}\right\rceil
    +\frac{N_bB}{2}(k_r-1)
    +2N_bB-3N_b+1~,\\
    Q_{\SEL}
    &=
    \frac{k_rN_bB}{2}
    +\left\lceil\log \left(\frac{N_bB}{2k_r}\right)\right\rceil
    +B~,
\end{align}
where $k_r$ is the QROAM parameter for loading the orbital-rotation data and $B$ is the bit precision of the Givens angles. The inverse preparation step contributes
\begin{equation}
    T_{\PREP^{\dg}} = \frac{N_b}{2k_P'}+k_P'+\mathfrak N+4n-3\eta+2b_r-9~,
\end{equation}
and $Q_{\PREP^\dg}=Q_{\PREP}$. The total cost of the walk operator is therefore
\begin{align}
    T({\hat{\mathcal{Q}}_W})
    &= \left\lceil\frac{N_b}{2k_P}\right\rceil
    + \left\lceil\frac{N_b}{2k_r}\right\rceil
    + \frac{N_b}{2k_P'}
    + (n+\mathfrak N+1)(k_P-1)\nonumber\\
    &\quad
    + \frac{N_bB}{2}(k_r-1)
    + k_P'
    + 2N_bB-3N_b
    + 2\mathfrak N
    + 8n
    -6\eta
    +4b_r
    -17~,\\
    Q({\hat{\mathcal{Q}}_W})
    &= N_b+2n+\mathfrak N+6+\frac{k_rN_bB}{2}
    +\left\lceil\log \left(\frac{N_bB}{2k_r}\right)\right\rceil
    +B~.
\end{align}

\begin{figure}[t!]
    \centering
    \begin{quantikz} 
    \lstick{succ $\ell$} & \gate[wires=2]{\text{prep}_{\ell}} & \qw & \ctrl{2} & \qw & \qw & \ctrl{5} & \qw & \qw & \qw & \gate[wires=2]{\text{prep}^{\dg}_{\ell}} & \qw \\ \lstick{$\ell$} & \qw & \gate{\text{In}_{\ell}} & \qw & \qw & \qw & \qw & \qw & \qw & \gate{\text{In}_{\ell}} & \qw & \qw \\ \lstick{\sign} & \qw & \gate[wires=2]{{\rm data}_{\ell}} \vqw{-1} & \gate{Z} & \qw & \qw & \qw & \qw & \qw & \gate[wires=2]{{\rm data}_{\ell}} \vqw{-1} & \qw & \qw \\ \lstick{rotations} & \qw & & \qw & \qw & \ctrl{2} & \qw & \ctrl{2} & \qw & & \qw & \qw \\ \lstick{spin} & \gate{H} & \qw & \ctrl{1} & \qw & \qw & \qw & \qw & \ctrl{1} & \qw & \gate{H} & \qw \\ \lstick{$|\psi_{\downarrow}\rangle$} & \qw & \qw & \swap{1} & \qw & \gate{R} & \gate{Z_1} & \gate{R^{\dg}} & \swap{1} & \qw & \qw & \qw \\ \lstick{$|\psi_{\uparrow}\rangle$} & \qw & \qw & \targX{} & \qw & \qw & \qw & \qw & \targX{} & \qw & \qw & \qw 
    \end{quantikz}
    \caption{\small Quantum circuit for the block encoding of the $\Gamma$-point density operator $\hat{\rho}_{\mbf q}\big|_{\Gamma}$. The circuit realizes the walk operator $\hat{\mathcal Q}_W=\PREP\cdot\SEL\cdot\PREP^\dg$. The $\PREP$ stage, represented by $\mathrm{prep}_\ell$, prepares the LCU index, sign, and alias-sampling registers. The $\SEL$ stage loads the Givens-rotation data through $\mathrm{In}_\ell$ and $\mathrm{data}_\ell$, applies the spin-controlled routing and the one-body transformation $(-1)^{s_p}\hat U_p Z_{p,\sigma}\hat U_p^\dg$ via the sequence of controlled SWAPs and $R\,Z_1\,R^\dg$, and then uncomputes the loaded data. The final application of $\mathrm{prep}_\ell^\dg$ disentangles the ancilla registers and completes the block encoding $U_{\hat{\rho}}$.\normalsize}
    \label{fig:rho_gamma_block_encoding_appendix}
\end{figure}
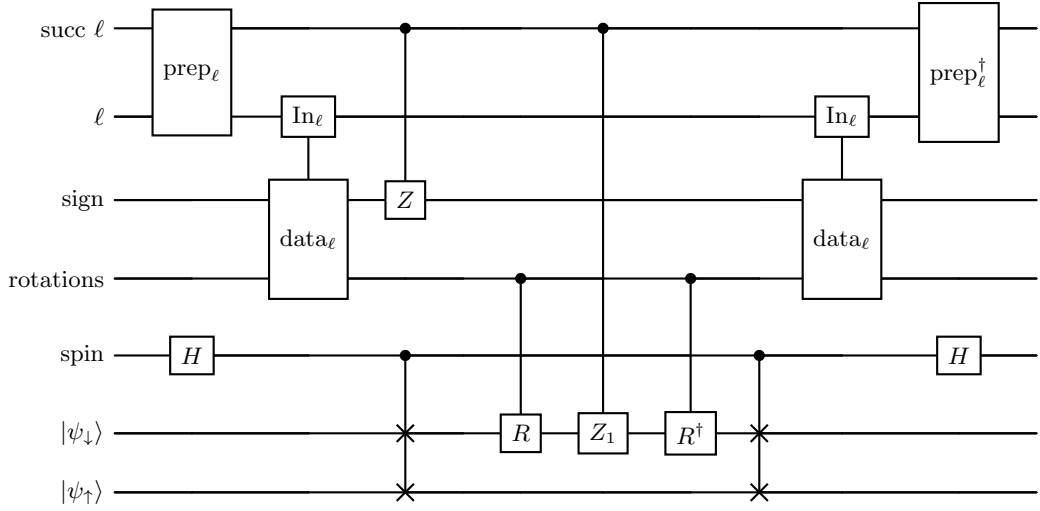
\vspace{5mm}
Since the density operator contains contributions from both signs and Hermitian components, the number of effective LCU entries doubles relative to the walk-operator construction. One therefore obtains
\begin{align}
    T({\hat{\rho}})
    &= \left\lceil\frac{N_b}{k_P}\right\rceil
    + \left\lceil\frac{N_b}{k_r}\right\rceil
    + \left\lceil\frac{N_b}{k_P'}\right\rceil
    + (n+\mathfrak N+2)(k_P-1)\nonumber\\
    &\quad
    + \frac{N_bB}{2}(k_r-1)
    + k_P'
    + 2N_bB-3N_b
    + 2\mathfrak N
    + 8n
    -6\eta
    +4b_r
    -9~,\\
    Q({\hat{\rho}})
    &= N_b+2n+\mathfrak N+6+\frac{k_rN_bB}{2}
    +\left\lceil\log \left(\frac{N_bB}{k_r}\right)\right\rceil
    +B~.
\end{align}
Fig. \ref{fig:rho_gamma_block_encoding_appendix} describes the corresponding circuit for the $\Gamma$-point density operator.
\subsection{Away from the \texorpdfstring{$\Gamma$}{Gamma}-point}

We now turn to the general density operator for fixed nonzero momentum transfer $\mbf q$. The corresponding walk operator is
\begin{equation}
    \hat{\mathcal{Q}}^{(\mbf{q})}_{W} = \PREP_{\mbf{q}}\cdot\SEL_{\mbf{q}}\cdot\PREP^{\dg}_{\mbf{q}}~.
\end{equation}
Relative to the $\Gamma$-point case, the key changes are that the LCU superposition must now include an explicit crystal-momentum label, and the select operation must route the controlled transformation to the pair of sectors $(\mbf k,\mbf k\oplus \mbf q)$. The $\PREP_{\mbf q}$ stage prepares
\begin{equation}
    \PREP_{\mbf{q}}\cdot|0^{n_{\ell}}\rangle|0\rangle|0\rangle
    =
    |L\rangle
    =
    \frac{1}{\sqrt{2\lambda_{O}}}
    \sum_{\sigma}\sum_{\mbf{k}}^{N_k}\sum_{p=1}^{N_b/2}
    \sqrt{|\mu_{\mbf{k},p}|}\;|p\rangle\;|\mbf{k}\rangle\;|\sigma\rangle\;|s_{\mbf{k},p}\rangle~,
\end{equation}
where the total number of modes is
\begin{equation}
    L_{\mbf{q}} = \sum_{\mbf{k}}^{N_k}\Xi_{\mbf{k}}^{(\mbf{q})}~,
\end{equation}
with $\Xi_{\mbf{k}}^{(\mbf{q})}$ the rank of the density operator in the corresponding $\mbf k$-sector, and $n_\ell=\log(N_bN_k)$ the size of the combined orbital-momentum index register. The resulting preparation cost is
\begin{align}
    T_{\PREP_{\mbf{q}}}
    &=
    \left\lceil\frac{L_{\mbf{q}}}{k_P}\right\rceil
    +(n_{\ell}+\mathfrak N+1)(k_P-1)
    +\mathfrak N
    +4n_{\ell}
    -3\eta
    +2b_r
    -9~,\\
    Q_{\PREP_{\mbf{q}}}
    &=2n_{\ell}+\mathfrak N+7~.
\end{align}
The $\SEL_{\mbf q}$ stage must now swap not only spin sectors but also momentum sectors. The relevant Hilbert space factorizes as
\begin{equation}
    \bigoplus_{\mbf{k}}\mathcal{H}_{\mbf{k}}^{\uparrow}\otimes\mathcal{H}_{\mbf{k}}^{\downarrow}~,
\end{equation}
and the unitary $\hat U_{\mbf q}(\mbf k)$ acts only within the selected pair $(\mbf k,\mbf k\oplus \mbf q)$ for fixed spin. This requires a classical-quantum adder to prepare $|\mbf k\oplus \mbf q\rangle$ and momentum-controlled swaps to bring the relevant sectors into the active position. Using the constructions of Refs.~\cite{Rubin_bloch_2023,Gidney_2025}, the swap and adder costs are
\begin{equation}
    T_{\rm SWAP_{\mbf{k}}} = N_bN_k+6n_k~,
    \qquad
    T_{A_{\mbf{q}}} = \log N_k~,
\end{equation}
where $n_k=\log N_k$. Combining these ingredients yields
\begin{align}
    T_{\SEL_{\mbf{q}}}
    &=
    8n_k
    +\left\lceil\frac{L_{\mbf{q}}}{k_r}\right\rceil
    +b_q(k_r-1)
    +N_bN_k(B-2)
    +N_bN_k
    +1~,\\
    Q_{\SEL_{\mbf{q}}}
    &=
    n_{\ell}+2n_k+k_rb_q+B+3~.
\end{align}
The inverse preparation contributes
\begin{equation}
    T_{\PREP^{\dg}_{\mbf{q}}}
    =
    \left\lceil\frac{L_{\mbf{q}}}{k_P'}\right\rceil
    +k_P'
    +\mathfrak N
    +4n_{\ell}
    -3\eta
    +2b_r
    -9~.
\end{equation}
Collecting terms gives the total cost of the walk operator
\begin{align}
    T(\hat{\mathcal{Q}}^{(\mbf{q})}_{W})
    &=
    \left\lceil\frac{L_{\mbf{q}}}{k_P}\right\rceil
    + \left\lceil\frac{L_{\mbf{q}}}{k_r}\right\rceil
    + \left\lceil\frac{L_{\mbf{q}}}{k_P'}\right\rceil
    + (n_{\ell}+\mathfrak N+1)(k_P-1)\nonumber\\
    &\quad
    + b_q(k_r-1)
    + k_P'
    + N_bN_k(B-1)
    + 8n_k
    + 2\mathfrak N
    + 8n_{\ell}
    - 6\eta
    + 4b_r
    - 17~,\\
    Q(\hat{\mathcal{Q}}^{(\mbf{q})}_{W})
    &= 3n_{\ell}+\mathfrak N+2n_k+k_rb_q+B+10~.
\end{align}
As in the $\Gamma$-point case, the density operator itself differs from the walk operator by an overall doubling of the number of effective LCU entries. This gives
\begin{align}
    T({\hat{\rho}}_{\mbf{q}})
    &= \left\lceil\frac{2L_{\mbf{q}}}{k_P}\right\rceil
    + \left\lceil\frac{2L_{\mbf{q}}}{k_r}\right\rceil
    + \left\lceil\frac{2L_{\mbf{q}}}{k_P'}\right\rceil
    + (n_{\ell}+\mathfrak N+2)(k_P-1)\nonumber\\
    &\quad
    + b_q(k_r-1)
    + k_P'
    + N_bN_k(B-1)
    + 8n_k
    + 2\mathfrak N
    + 8n_{\ell}
    - 6\eta
    + 4b_r
    - 9~,\\
    Q({\hat{\rho}}_{\mbf{q}})
    &= N_bN_k+3n_{\ell}+\mathfrak N+2n_k+k_rb_q+B+13~.
\end{align}
\section{A note on the validity of the IR ansatz for static structure factor}
\label{app: note on the IR ansatz}
\begin{figure}[ht!]
    \centering
    \includegraphics[width=0.75\linewidth]{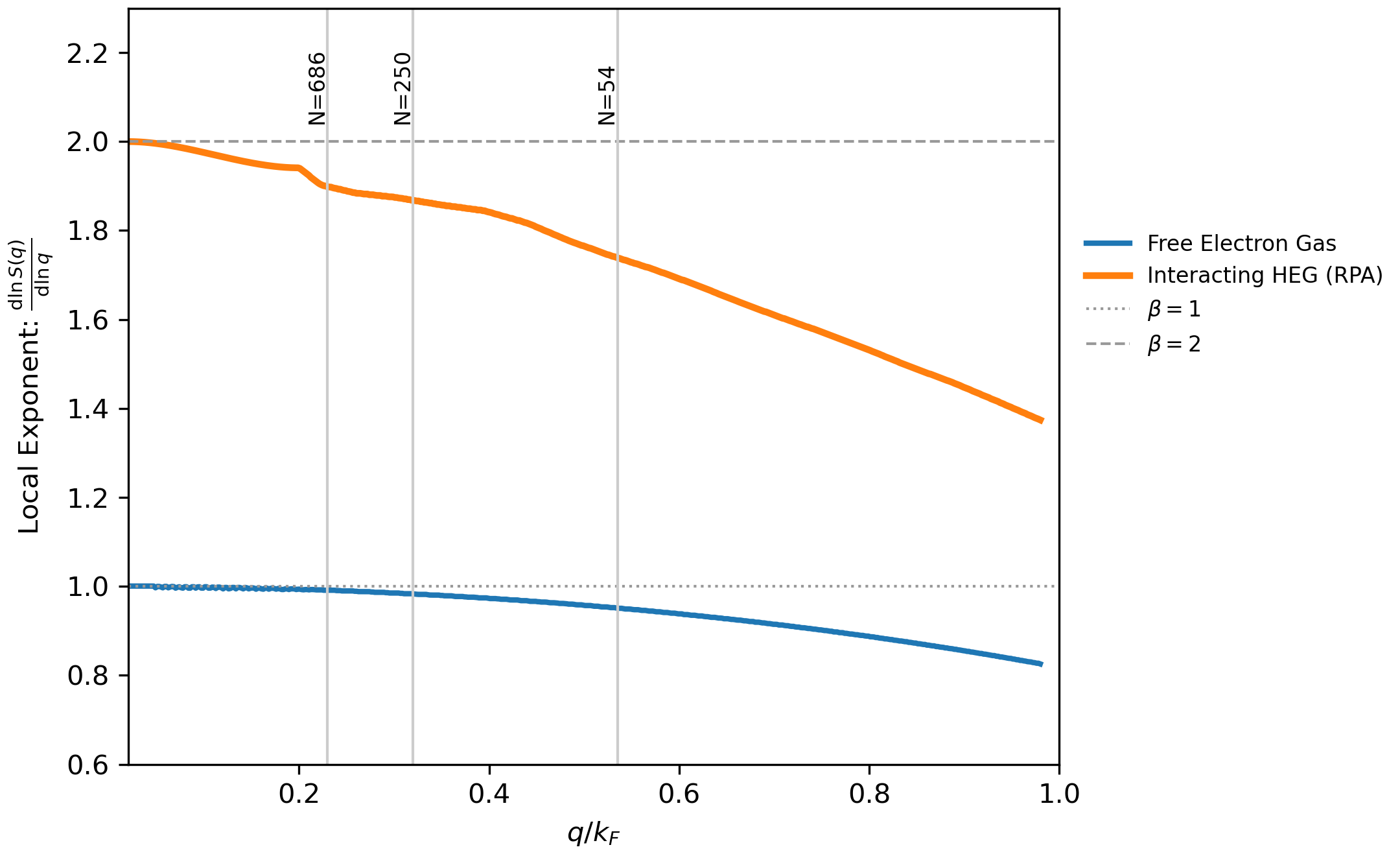}
    \caption{Local logarithmic exponent of the static structure factor, $\beta(q)=\frac{ d \ln S(q)}{d \ln q}$, for the three-dimensional homogeneous electron gas. The free-electron gas approaches the expected low-momentum linear behavior, $\beta=1$, while the interacting Coulomb gas approaches the plasmon-controlled quadratic limit, $\beta=2$, as $q/k_F \to 0$. Vertical lines mark the smallest available reciprocal-lattice shell for representative finite electron counts $N=$ 686, 250, and 54, illustrating how typical finite cells sample the crossover rather than the strict asymptotic regime.}
    \label{fig:local_exponent}
\end{figure}
Anisotropic materials or metals with complex multi-sheet Fermi surfaces may fall outside the scope of our $S(q) \sim \alpha q^\beta$ algebraic ansatz. We can, however, still treat these materials within our approach by partitioning the static structure factor into a mean-field component and a quantum correction from the FTQC to separate the macroscopic dielectric response from the localized many-body correlation.

We can achieve this separation by relying on the (classical) random phase approximation (RPA) to compute the exact long-wavelength continuum. We can construct the classical $S_{\text{RPA}}(q)$ from the linear density response of a non-interacting homogeneous electron gas to an external scalar potential, the Lindhard response \cite{Lindhard_1954}:

\begin{equation}
\label{eqn:chi0}
\chi_0(q, \omega) = 2 \int \frac{d^3k}{(2\pi)^3} \frac{f(\epsilon_k) - f(\epsilon_{k+q})}{\hbar\omega + \epsilon_k - \epsilon_{k+q} + i\eta}
\end{equation}

\noindent where $f(\epsilon)$ is the Fermi-Dirac distribution, $\epsilon_k$ is the single-particle band dispersion, $\hbar\omega$ and $q$ are the energy and momentum transfers, and $\eta \to 0^+$ is a positive infinitesimal parameter (for causaility). We can now insert this susceptibility into the zero-temperature fluctuation-dissipation theorem to obtain the classical structure factor as:

\begin{equation}
\label{eqn:SRPA}
S_{\text{RPA}}(q) = -\frac{1}{\pi n} \int_0^\infty d\omega \, \text{Im} \left[ \frac{\chi_0^R(q, \omega)}{1 - v(q)\chi_0^R(q, \omega)} \right]
\end{equation}

\noindent where $n$ is the average electron density, $\chi_0^R(q, \omega)$ is the retarded Lindhard susceptibility, and $v(q) = 4\pi/q^2$ is the bare Coulomb potential.

The denominator in Equation~\ref{eqn:SRPA} defines the RPA dielectric function, $\epsilon_{\text{RPA}}(q, \omega) = 1 - v(q)\chi_0^R(q, \omega)$. This term represents an infinite resummation of bubble (or ring) diagrams, which captures the long-range Coulomb screening and collective plasmon modes. From this, we can see that $S_{\text{RPA}}(q)$ obeys the correct asymptotic physics in the $q \to 0$ limit. However, because RPA neglects vertex corrections, it fails to capture short-range exchange and local many-body correlation effects; it misses the exchange-correlation hole.

Figure~\ref{fig:local_exponent} illustrates why this issue can arise even in the comparatively simple homogeneous electron gas. The local logarithmic exponent $\beta(q)=d\ln S(q)/d\ln q$ approaches its asymptotic value only in the strict $q/k_F\to 0$ limit, while the smallest momentum accessible in representative finite cells can lie in a broad pre-asymptotic crossover region. Thus, a finite calculation may not directly sample the regime where a single power law $S(q)\sim \alpha q^\beta$ is quantitatively reliable. We can recover these missing many-body effects using our FTQC routine to evaluate the exact structure factor on a discrete mesh and write the residual correlation difference as $\Delta S(q) = S_{\text{FTQC}}(q) - S_{\text{RPA}}(q)$. By subtracting the mean-field $S_{\text{RPA}}(q)$, this $\Delta S(q)$ isolates the exchange-correlation hole missing from the RPA description. Provided the material does not exhibit competing emergent long-range orderings, for instance, charge or spin density waves that could introduce singularities at specific $q$-vectors, the correlation hole is highly localized in real space. Thus, its momentum-space Fourier transform is smooth and slowly varying. This allows us to fit the smooth residual points to a continuous curve and add it back to the exact classical continuum to yield a thermodynamic-limit surrogate, $S_{\text{surrogate}}(q) = S_{\text{RPA}}(q) + \Delta S(q)$, without the need for large supercells to capture small $q$ on the FTQC.

The challenge is to quantify how big a supercell one needs. We don't know {\it a priori} where $S_{\text{RPA}}(q)$ becomes a poor approximation. One thing we could consider is using a better classical compute approximation than RPA (like QMC) and see where the solutions start to diverge. 
\bibliography{main}
\end{document}